\documentclass[preprint,12pt]{elsarticle}

\usepackage{amssymb}
\usepackage{amsmath}

\usepackage[version=4]{mhchem}
\usepackage[per-mode=symbol]{siunitx}
\usepackage{dirtytalk}

\newcommand{\dmm}[1]{{\color{black}{#1}}}
\newcommand{\lpp}[1]{{\color{black}{#1}}}
\newcommand{\abk}[1]{{\color{black}{#1}}}

\newcommand{\ajk}[1]{{\color{black}{#1}}}
\newcommand{\lppr}[1]{{\color{black}{#1}}}

\journal{Combustion and Flame}

\begin{document}

\begin{frontmatter}



\title{On the spatial structure and intermittency of soot in a lab-scale gas \dmm{turbine} combustor: insights from large-eddy simulations}


\author[affi1]{Leonardo Pachano\corref{cor1}}\ead{leonardo.pachano@bsc.es} 
\author[affi1]{Daniel Mira} 
\author[affi2]{Abhijit Kalbhor}
\author[affi2,affi3]{Jeroen van Oijen} 

\affiliation[affi1]{organization={Barcelona Supercomputing Center (BSC)},
            addressline={Plaça d'Eusebi Güell 1-3}, 
            city={Barcelona},
            postcode={08034}, 
            country={Spain}}

\affiliation[affi2]{organization={Department of Mechanical Engineering, Eindhoven University of Technology},
            city={Eindhoven},
            postcode={5600 MB},
            country={the Netherlands}}

\affiliation[affi3]{organization={Eindhoven Institute for Renewable Energy Systems (EIRES)},
            city={Eindhoven},
            postcode={5600 MB},
            country={the Netherlands}}

\cortext[cor1]{Corresponding author}

\begin{abstract}

\dmm{This work presents a numerical investigation of soot formation in the Cambridge lab-scale gas turbine combustor. Large-eddy simulations (LES) of a swirl-stabilized ethylene flame are performed using the flamelet generated manifold method coupled with a discrete sectional model to account for soot formation, growth, and oxidation. The study aims to elucidate the mechanism governing the spatial structure and intermittency of soot, supported by comparisons with experimental data. The predicted soot distribution agrees well with measurements, with peak concentrations near the bluff body. Flow recirculation is identified as the key mechanism driving soot accumulation in fuel-rich regions, where surface reactions dominate soot mass growth. \lpp{Soot intermittency arises from fluctuations in the flow field driven by interactions between the flame front and the recirculation vortex}. \lpp{Two soot modeling approaches are evaluated, differing in \abk{their treatment of} soot model quantities: \abk{the first approach} employs on-the-fly computation of source terms (FGM-C), while the \abk{second} uses fully pre-tabulated source terms (FGM-T).} Their predictive performance and computational cost are compared in the context of 
unsteady, sooting flames in swirl-stabilized combustors.}

\end{abstract}




\end{frontmatter}




\section*{Novelty and Significance Statement}

This paper offers detailed insights into soot production within a swirl-stabilized gas turbine combustor by employing a framework that couples Flamelet Generated Manifold chemistry with a Discrete Sectional Method for soot formation. Notably, this study provides a detailed characterization of flow-soot interactions, revealing the mechanism that governs the spatial structure of soot within the Cambridge rich-quench-lean combustor. Furthermore, it presents the first numerical investigation of soot intermittency in this turbulent flame configuration. \lppr{This study also constitutes the first direct evaluation of the FGM-T formulation with soot section clustering in comparison with the more detailed FGM-C formulation within a lab-scale gas turbine combustor, offering insight into their respective predictive performance and computational cost.}

\section*{Author Contributions}

\textbf{LP:} Conceptualization, Formal analysis, Visualization, Writing – original draft. \textbf{DM:} Conceptualization, Supervision, Software, Resources, Writing – original draft. \textbf{AK:} Methodology, Software, Writing – review \& editing. \textbf{JvO:} Conceptualization, Methodology, Writing – review \& editing. 


\section{Introduction}
\label{sec:introduction}

Soot emissions from combustion systems pose significant concerns due to their harmful impact on the environment and their contribution to respiratory and cardiovascular diseases. In the aviation sector, soot production in gas turbine engines is of particular relevance. Soot emissions from aircraft engines contribute to air pollution in the vicinity of airports \cite{alzahrani2024} and at high altitudes, where they can affect cloud formation and radiative forcing \cite{li2024}. The design of low-emission combustors and the implementation of alternative fuels rely heavily on the predictive capability of numerical models to simulate soot production. Understanding and accurately modeling soot formation processes is therefore crucial for the development of cleaner combustion technologies.

\dmm{A fundamental challenge in developing accurate soot models lies in the complex, multi-scale nature of soot formation, which involves tightly coupled interactions of physical and chemical processes, including fuel pyrolysis, polycyclic aromatic hydrocarbon (PAH) growth, \lpp{soot} nucleation, surface growth, coagulation, and oxidation~\cite{thomson2023}. These processes are highly sensitive to local thermochemical conditions, such as temperature, pressure, fuel composition, and turbulence-chemistry interactions. Capturing this complexity in predictive numerical models requires rigorous validation against high-quality experimental data. To this end, several well-characterized lab-scale gas turbine combustor configurations have been 
developed in the literature.}

Two well-established configuration are the rich-quench-lean (RQL) combustors developed by the German Aerospace Center (DLR) \cite{Geigle.2015} and the University of Cambridge \cite{elhelou2021,defalco2021}. \dmm{These benchmark cases provide access to detailed measurements of flow fields, temperature, species concentrations, and soot properties under controlled conditions, thereby enabling systematic comparisons with high-fidelity simulations and supporting the advancement of physically grounded soot model development.}

\lppr{While both the DLR and Cambridge combustors represent relevant RQL-type architectures, their distinct operating conditions and fuel–air injection strategies lead to fundamentally different soot-formation pathways. The DLR configuration operates under globally rich ($\phi=1.2$) and pressurized (3–\SI{5}{bar}) conditions, which promote high soot loadings \cite{Geigle.2015}. In contrast, the Cambridge combustor operates under globally lean ($\phi=0.3$) and atmospheric pressure conditions, allowing for the analysis of soot formation in the limit of low concentrations, representative of lean-burn engines. Furthermore, the DLR combustor's annular fuel-injection strategy creates isolated, fuel-rich shear-layer pockets that drive soot formation \cite{chong2018,garcia-oliver2024}, resulting in a comparatively uniform soot distribution along the rich branches. Conversely, the Cambridge burner’s jet-in-swirling-flow configuration leads to a distinct soot spatial structure characterized by a maximum concentration near the bluff body followed by a progressive downstream decay \cite{elhelou2021,defalco2021}.}

A considerable body of work has focused on soot modeling in the DLR combustor using Large-Eddy Simulations (LES). Different levels of modeling complexity have been explored, ranging from two-equation soot models \cite{felden2018, eberle2018} to more advanced formulations such as three-equation models \cite{franzelli2023}, the method of moments \cite{chong2018,Cokuslu.2022}, and sectional methods \cite{Grader.2018,garcia-oliver2024}. These studies have provided valuable insights into the influence of turbulence-chemistry interaction, local thermochemical conditions, and soot model formulations on soot evolution. In contrast, LES studies of soot production in the Cambridge combustor are comparatively less extensive. Existing numerical investigations have employed two-equation models \cite{Giusti.2018} and sectional methods \cite{gkantonas2020,pachano2024}. These studies have highlighted the intricate interplay between soot formation and the underlying mixing and combustion processes within this specific combustor geometry.

The present work contributes to ongoing advancements in soot modeling by employing a Discrete Sectional Method (DSM) integrated within the Flamelet Generated Manifold (FGM) chemistry framework, referred to as FGM-DSM. The fundamental principles and validation of the FGM-DSM approach were initially demonstrated in laminar flame studies \cite{kalbhor2021}, and later extended to \lpp{LES} for turbulent non-premixed jet flames (LES-FGM-DSM) \cite{kalbhor2024}. 
The LES-FGM-DSM framework was developed in two distinct formulations: (i)~a runtime approach where soot source terms are dynamically computed based on local tabulated thermochemical states (FGM-C), and (ii)~a fully pre-tabulated approach where soot source terms are stored in the manifold and retrieved \lpp{at} runtime (FGM-T). From the FGM-T formulation, a computationally-efficient variant was developed based on clustering of soot sections. This approach, termed FGM-CDSM, was first validated in laminar flame configurations in steady and transient conditions~\cite{kalbhor2023}.
The LES extension of this clustered formulation (LES-FGM-CDSM) has recently been applied to the DLR burner \cite{garcia-oliver2024} and the Cambridge combustor \cite{pachano2024}. In this formulation, soot evolution is computed assuming a pre-defined range of particle sizes within each cluster, which has been shown to adequately capture the spatial distribution and magnitude of soot in these complex geometries. 

\lppr{While previous works on the Cambridge burner have investigated soot predictions under various operating conditions~\cite{Giusti.2018,gkantonas2020,pachano2024}, important gaps remain regarding the mechanisms governing the spatial soot structure. In particular, soot intermittency has not yet been examined numerically in this configuration. Moreover, although the FGM-T formulation with soot section clustering has been applied recently~\cite{garcia-oliver2024,pachano2024}, a direct comparison with the FGM-C formulation in a lab-scale gas turbine combustor is still lacking. To address these limitations, the present study conducts a fundamental analysis of soot formation and dynamics in the Cambridge combustor. Physically, the work aims to elucidate the underlying mechanisms of soot distribution, with particular emphasis on spatial structure, intermittency, and the dominant physical drivers in this swirl-stabilized configuration. From a modeling perspective, the study evaluates the predictive capabilities of the fully tabulated approach by directly comparing the FGM-C and FGM-T formulations against experimental measurements.}

\lpp{The rest of the paper is organized as follows.} Details on the modeling approach are provided in Section~\ref{sec:modeling}. Section~\ref{sec:case_study} presents the studied configuration, describing the geometry of the combustor and the numerical setup employed for the \dmm{LES.} Section~\ref{sec:results} introduces soot results, examining \dmm{the spatial distribution, the evolution of particle size distributions (PSD), the interaction between flow and soot particles, and soot intermittency.} Finally, Section~\ref{sec:conclusions} summarizes the main findings and provides concluding remarks.

\section{Combustion and soot modeling approach}
\label{sec:modeling}

\dmm{The modeling approach in this study adopts an integrated modeling framework to simulate turbulent combustion with soot formation in the context of LES. The \lpp{FGM} approach~\cite{vanoijen2016} is used to describe the chemical kinetics characteristic of combustion, while \lpp{a DSM}~\cite{Hoerlle.2019} is employed to evaluate the detailed evolution of soot particles. The resulting modeling approach (FGM-DSM) is extended to LES by using a presumed-shape \lpp{probability density function} (PDF) for the control variables to account for sub-filter turbulent chemistry interactions~\cite{kalbhor2021,kalbhor2023,kalbhor2024,garcia-oliver2024,pachano2024} and it is referred \lpp{to} as LES-FGM-DSM.
\lppr{The manifold space is parametrized through mixture fraction ($Z$) and its variance ($Z_v$), a chemical progress variable ($Y_c$), and an scaled enthalpy ($\mathcal{H}$) to account for heat losses effects.} The specific parameter settings employed 
for this study are presented in the following sections.}

The governing equations for LES are derived from the filtered Navier-Stokes equations under the assumption of low-Mach number flows and unity Lewis number \cite{both2020}. The filtered thermo-fluid dynamic equations, representing the conservation of continuity, momentum, and \dmm{enthalpy are given by:}
\begin{align}
    \label{eqn:eq_continuity}
    \frac{\partial \bar{\rho}}{\partial t} + \boldsymbol{\nabla} \cdot \left( \bar{\rho} \boldsymbol{\tilde{u}} \right) &=0,\\
    \label{eqn:eq_momentum}
    \frac{\partial \bar{\rho} \boldsymbol{\tilde{u}}}{\partial t} + \boldsymbol{\nabla} \cdot \left( \bar{\rho} \boldsymbol{\tilde{u}} \boldsymbol{\tilde{u}} \right) &=
    - \boldsymbol{\nabla}  \cdot \overline{\tau}_M -\boldsymbol{\nabla} \bar{p} +\boldsymbol{\nabla}  \cdot \left( \bar{\mu} \boldsymbol{\nabla}  \boldsymbol{\tilde{u}} \right),\\
     \label{eqn:eq_enthalpy}
    \frac{\partial \overline{\rho} \tilde{h}}{\partial t} + \boldsymbol{\nabla} \cdot \left( \overline{\rho} \boldsymbol{\tilde{u}} \tilde{h} \right) &=
    - \boldsymbol{\nabla}  \cdot \overline{\tau}_{h} + \boldsymbol{\nabla}  \cdot \left( \bar{\rho}  \bar{D} \boldsymbol{\nabla}  \tilde{h} \right).
\end{align}

In Eq.~\eqref{eqn:eq_continuity}, Eq.~\eqref{eqn:eq_momentum}, and Eq.~\eqref{eqn:eq_enthalpy}, $\bar{\rho}$ denotes density, $\tilde{\boldsymbol{u}}$ velocity, $\bar{p}$ pressure, $\tilde{h}$ enthalpy (encompassing both sensible and chemical contributions), $\tilde{D}$ diffusivity, and $\overline{\tau}$ unresolved fluxes. \dmm{Note that} \lpp{an} 
overbar indicates Reynolds-filtered quantities, \dmm{while tilde refers to} Favre-filtered quantities. 

To close the system of equations, models are required for the unresolved heat and momentum fluxes. Here, a gradient diffusion hypothesis~\cite{miramartinez2014} is used to model the unresolved heat fluxes, and the Boussinesq approximation~\cite{poinsot2005} is employed for the unresolved momentum transport. The subgrid-scale eddy viscosity ($\nu_t$) required in these models is estimated using Vreman’s model~\cite{vreman2004}, with the model constant set to $c_k = 0.1$. This value is consistent with recommendations and findings from prior research in similar applications \cite{mira2020,kalbhor2024}.

\subsection{FGM chemistry}
\label{subsec:FGM_model}

The tabulated chemistry employed in this study is based on a laminar flamelet database, 
constructed by \dmm{a set of} one-dimensional counterflow diffusion flames 
\dmm{at different strain rates}
using the code CHEM1D \cite{somers1994}. 
The gas-phase chemistry within the flamelet framework 
is described using the detailed KM2 chemical kinetics scheme 
developed by Wang et al. \cite{Wang.2013}. \dmm{The laminar database is constructed
by a series of flamelets across the stable branch of the S-curve until the extinction point is \lpp{reached},
while the rest of the manifold, derived from the extinction limit, is covered by an
extinguishing flamelet. A parametrization in terms of $Z$, 
following Bilger’s formulation~\cite{bilger1989}, and $Y_c$ is employed. To account for heat loss, 
a radiative source term is introduced in the enthalpy equation to obtain the
flamelet solutions with different enthalpy levels, so a three-dimensional space including the scaled enthalpy $\mathcal{H}$ is used to describe the manifold space~\cite{kalbhor2024,garcia-oliver2024}.} 
$Y_c$ is defined as a linear combination of the mass 
fractions of key combustion species: water vapor (\ce{H2O}), carbon monoxide 
(\ce{CO}), molecular hydrogen (\ce{H2}), carbon dioxide (\ce{CO2}), acetylene 
(\ce{C2H2}), and pyrene (\ce{A}4). $\mathcal{H}$ 
is normalized between the minimum and maximum enthalpy levels observed 
in the flamelet for a given mixture fraction, and is mathematically 
expressed as $\mathcal{H} = \left(h - h_{min}\right)/\left(h_{max}- h_{min}\right)$.
 
To account for the interaction between turbulence and chemistry \dmm{in the sub-filter scale}, a presumed \lpp{PDF} approach is employed. Specifically, a $\beta$-PDF function is employed to represent the statistical distribution of the mixture fraction. This choice is supported by previous LES studies that have demonstrated the suitability of the $\beta$-PDF for flamelet-based combustion modeling~\cite{domingo2008, massey2021,kalbhor2024}. The variance of the mixture fraction, is calculated as \lppr{$\tilde{Z}_v = \tilde{Z^2}-\tilde{Z}\tilde{Z}$}. In contrast to the mixture fraction, a $\delta$-function is used to represent the probability distributions of both $Y_c$ and $\mathcal{H}$ \cite{govert2015,govert2018}.

During the simulation runtime, the filtered transport equations for the \dmm{selected} control variables are solved to reconstruct the thermochemical state at each computational cell. The governing equations for the mixture fraction, the reaction progress variable, and the mixture fraction variance are \dmm{given by Eq.~\eqref{eqn:eq_control_Yc}, Eq.~\eqref{eqn:eq_control_Z}, and \lppr{Eq.~\eqref{eqn:eq_control_Zv}}:}
\begin{align}
    \label{eqn:eq_control_Z}
    \frac{\partial \overline{\rho} \tilde{Z}}{\partial t} + \boldsymbol{\nabla} \cdot \left( \overline{\rho} \boldsymbol{\tilde{u}} \tilde{Z} \right) &=
    - \boldsymbol{\nabla}  \cdot \overline{\tau}_{Z} +  
    \boldsymbol{\nabla}  \cdot \left( \bar{\rho}  \bar{D} \boldsymbol{\nabla}  \tilde{Z} \right),\\
    \label{eqn:eq_control_Yc}
    \frac{\partial \overline{\rho} \tilde{Y_c}}{\partial t} + \boldsymbol{\nabla} \cdot \left( \overline{\rho} \boldsymbol{\tilde{u}} \tilde{Y_c} \right) &=
    -\boldsymbol{\nabla}  \cdot \overline{\tau}_{Y_c} + \boldsymbol{\nabla}  \cdot \left( \bar{\rho}  \bar{D} \boldsymbol{\nabla}  \tilde{Y_c} \right) +  \overline{\dot\omega}_{Y_c},\\
    \label{eqn:eq_control_Zv}
    \frac{\partial \overline{\rho} \lppr{\tilde{Z}_v}}{\partial t} + \boldsymbol{\nabla} \cdot \left( \overline{\rho} \boldsymbol{\tilde{u}} \lppr{\tilde{Z}_v} \right) &= - \boldsymbol{\nabla}  \cdot \overline{\tau}_{Z_v} + \boldsymbol{\nabla}  \cdot \left( \bar{\rho}  \bar{D} \boldsymbol{\nabla} \lppr{\tilde{Z}_v} \right) \\\nonumber
    & - 2 \overline{\tau}_{Z} \cdot \boldsymbol{\nabla}  \tilde{Z} -2 \overline{s}_{\chi_Z}.
\end{align}

In Eq.~\eqref{eqn:eq_control_Yc}, the term \dmm{$\bar{\dot{\omega}}_{Y_c}$} represents the filtered chemical source term for the reaction progress variable. The unresolved transport of scalars, denoted as $\bar{\tau}_{Z}$ in Eq.~\eqref{eqn:eq_control_Z} and $\bar{\tau}_{Y_c}$ in Eq.~\eqref{eqn:eq_control_Yc}, arises from the filtering process applied to the governing equations. This unresolved transport is modeled using a gradient diffusion approach \cite{miramartinez2014}. Similarly, the unresolved contribution to the scalar dissipation rate for the mixture fraction, denoted by $\bar{S}_{\chi_Z}$, is modeled using a linear relaxation assumption at the \dmm{sub-filter} scale~\cite{domingo2008}. This model expresses $\bar{S}_{\chi_Z}$ as $\bar{\rho} \lppr{\tilde{Z}_v} \nu_{t} / \Delta^2$, where $\Delta$ is the filter width associated with the computational mesh.

To efficiently access the thermochemical data stored in the flamelet manifold, a normalized progress variable ($C$) is employed. Subsequently, any tabulated quantity is retrieved from the database using Eq.~\eqref{eqn:eq_PDF}:
\dmm{
\begin{align}
    \label{eqn:eq_PDF}
    \tilde{\psi} \left(C, Z, \mathcal{H}\right) = \int_{0}^{1} \int_{0}^{1} \int_{0}^{1} \psi \left(C, Z, \mathcal{H}\right) \tilde{P}\left(C, Z, \mathcal{H}\right) dC dZ d\mathcal{H},
\end{align}

where $\tilde{P}\left(C, Z, \mathcal{H}\right)$ represents the joint PDF of C, Z, and $\mathcal{H}$~\cite{ihme2008}. However, as the sub-filter fluctuations of the progress variable were introduced in the sub-filter mixture fraction variance and the heat-loss PDF is assumed to follow a $\delta$-function, the manifold is represented by Eq.~\eqref{eqn:eq_PDF_aprox}:

\begin{align}
    \label{eqn:eq_PDF_aprox}
    \tilde{\psi} \left(C, Z, \mathcal{H}\right) \approx \int_{0}^{1} \psi \left(Z \right) \tilde{P}\left( Z\right) dZ.
\end{align}
}

The turbulent flamelet database itself is discretized using a four-dimensional grid with dimensions of $101\times11\times101\times21$ points, corresponding to the $Z \times Z_v \times Y_c \times \mathcal{H}$ space, respectively.

\subsection{The sectional soot model}
\label{subsec:soot_model}

Following the \lpp{DSM} formulation, the population of soot particles is discretized into \lpp{an} specified number of sections \dmm{based on particle volume, \lpp{in this case} $n_{sec}=30$}. At the flamelet level, the soot mass fraction ($Y_{s,i}$) within each section $i$ is governed by a transport equation (Eq.~\eqref{eqn:soot_flamelet}) that considers several key processes: 
\dmm{advection,} thermophoresis, diffusion, and soot kinetics. In this equation, $\rho$ represents the density, $\boldsymbol{u}$ the velocity, $\nu_T$ the thermophoretic velocity (calculated using the expression proposed by Frienlander \cite{friendlander2000smoke}), $D_{s}$ the soot diffusion coefficient (assumed constant across all sections for numerical \dmm{stability~\cite{kalbhor2021})}, and $\dot{\omega}_{s,i}$ the chemical source term for section $i$. \lpp{The transport equation for $Y_{s,i}$} \dmm{is given by:

\begin{align}
    \begin{split}
    \label{eqn:soot_flamelet}
    \frac{\partial \left( \rho Y_{s,i} \right)}{\partial t} + {}\nabla \cdot \left(\rho \left[ \boldsymbol{u} + \nu_T \right]Y_{s,i} \right) =
    \nabla \cdot \left(\rho D_{s} \nabla Y_{s,i} \right) + \dot{\omega}_{s,i}, \,
    \forall \ i \in \left[1, n_{sec} \right]
    \end{split}
\end{align}
}

The soot chemistry model incorporates \dmm{several pathways} for soot formation and destruction. These include nucleation, modeled based on the dimerization of \ce{A}4 molecules; PAH condensation, which accounts for collisions between soot particles and \ce{A}4 molecules; coagulation, modeled according to the work of Kumar et al. \cite{kumar1996}; and surface chemistry, encompassing particle growth and oxidation through the \ce{H}-Abstraction \ce{C2H2}-Addition (HACA) mechanism \cite{Frenklach.1991,Appel.2000}. Within the HACA mechanism, oxidation considers the interaction of soot particles with both \ce{O2} and \ce{OH}, with a \ajk{commonly adopted} collision efficiency ($\gamma_{\ce{OH}}$) of 0.13 for the latter~\cite{Neoh.1985,Frenklach.1991}. For simplicity \dmm{on DSM tabulation~\cite{kalbhor2020,kalbhor2023}}, soot particles are treated as spheres, neglecting potential morphological complexities. The consumption of soot precursor species is accounted for during the one-dimensional flamelet calculations through a two-way coupling between the gas and solid phases. A more detailed description of the soot model and its underlying sub-processes, along with validation results, can be found in Hoerlle et al. \cite{Hoerlle.2019}.

In the LES context, the filtered transport equation for the soot mass fraction in section $i$ is given by Eq.~\eqref{eq:soot_alya}:
\dmm{
\begin{equation}     \label{eq:soot_alya}
    \frac{\partial (\bar{\rho} \widetilde{Y}_{s,i}) }{\partial t} + \nabla \cdot \left(\bar{\rho} \left[ \boldsymbol{\widetilde{u}} + \boldsymbol{\widetilde{v}}_T \right] \widetilde{Y}_{s,i} \right) =
    \nabla \cdot \left (\bar{\rho} D_{T} \nabla \widetilde{Y}_{s,i} \right)
    + \overline{\dot{\omega}}_{s,i}, \forall \ i \in \left[1, n_{sec} \right],
\end{equation}
where $D_{T}$ accounts for turbulent diffusion modeled by the gradient diffusion hypothesis~\cite{kalbhor2023}. Note} this equation includes a filtered thermophoretic velocity, defined as $\boldsymbol{\widetilde{v}}_T\!=\!-0.554 \bar{\nu} (\nabla \widetilde{T} / \widetilde{T})$, and a filtered source term for section $i$, denoted as $\overline{\dot{\omega}}_{s,i}$.

\subsubsection{Runtime computation of soot source terms (FGM-C)}
\label{subsubsec:FGM-C_model}

The \dmm{FGM-C approach relies on the computation of the 
soot source terms at runtime. It requires the} tabulated filtered thermochemical parameters and the local concentrations of gas-phase species that are relevant to the soot model.
The filtered soot source term is approximated by neglecting the \dmm{contribution associated to the interaction between soot and turbulence at the sub-filter scale}, as shown in Eq.~\eqref{eqn:omegaYs_FGM-C}:
\begin{align}
    \label{eqn:omegaYs_FGM-C}
    \overline{\dot{\omega}}_{s,i} = \dot{\omega}_{s,i}\left(\widetilde{\phi}_{g},\widetilde{\phi}_{s}\right),
\end{align}
\noindent where $\widetilde{\phi}_{g}$ and $\widetilde{\phi}_{s}$ represent filtered gas-phase thermochemical and soot variables, respectively. 
  
\subsubsection{Tabulation of soot source terms with section clustering (FGM-T)}
\label{subsubsec:FGM-T_model}

In contrast to the FGM-C approach, \dmm{the tabulated version, FGM-T, is based on the
pre-calculation of the soot source terms during flamelet generation and their tabulation in the chemical manifold. The FGM-T \lpp{approach} can be extended to the clustered \lpp{formulation} to reduce the computational cost without \lpp{compromising} accuracy~\cite{kalbhor2023,kalbhor2024,pachano2024,garcia-oliver2024}, so} the original thirty soot sections are grouped into a smaller number of clusters, $n_{clt}=6$, based on the assumption that the soot PSD within each cluster remains consistent with the original sections. The soot mass fraction for a given cluster $j$, denoted as $Y_{s,j}$, is then calculated by summing the soot mass fractions of the individual sections $i$ that belong to that cluster: $Y_{s,j} = \sum_{i=i_{j}^\mathrm{min}}^{i_{j}^\mathrm{max}} Y_{s,i}$, where $i_{j}^\mathrm{min}$ and $i_{j}^\mathrm{max}$ are the lower and upper limits of the section indices within cluster $j$, respectively. By reducing the number of sections to six clusters, the \dmm{computational cost} in the LES is significantly lowered, as only six transport equations are required to model soot formation. Following the presumed PDF approach used for the gas phase, the filtered source term for each cluster $j$, $\overline{\dot{\omega}}_{s,j}$, is determined using Eq.~\eqref{eqn:soot_pdf}:
\begin{align}
    \begin{split}
    \label{eqn:soot_pdf}
    \overline{\dot{\omega}}_{s,j} = \bar{\rho} \iint\frac{1}{\rho} \dot{\omega}_{s,j} 
    \widetilde{P}\left(\phi_g,\phi_s\right) d\phi_g d\phi_s.
    \end{split}
\end{align}

In Eq.~\eqref{eqn:soot_pdf}, the joint PDF is approximated as $\widetilde{P}(\phi_g,\phi_s)\!=\!\widetilde{P}(\phi_g)P(\phi_s|\phi_g)$. Here, the marginal PDF for the gas phase, $\widetilde{P}(\phi_g)$, is represented by a $\beta$-function for the mixture fraction, and the conditional PDF for the soot phase given the gas phase, $P(\phi_s|\phi_g)$, is represented by a $\delta$-function. This approximation allows for a partial accounting of turbulence-soot interactions through the \dmm{sub-filter} fluctuations of the mixture fraction. Similar \ajk{to treatment of slow–time-scale chemistry within the flamelet framework~\cite{ihme2008},} the soot cluster source term is separated into production and consumption terms: $\overline{\dot{\omega}}_{s,j} = \overline{\dot{\omega}}_{s,j}^{+} + \overline{\dot{\omega}}_{s,j}^{-}$.  These terms are then approximated according to Eq.~\eqref{eqn:omegaYs_FGM-T}, where the superscript \say{tab} indicates tabulated quantities. 

\begin{align}
    \label{eqn:omegaYs_FGM-T}
    \overline{\dot{\omega}}_{s,j} \approx \left[\overline{\dot{\omega}}_{s,j}^{+} \right]^\mathrm{tab}+\widetilde{Y}_{s,j} \left[ \overline{\frac{\dot{\omega}_{s,j}^{-}}{Y_{s,j}}} \right]^\mathrm{tab}
\end{align}

This approximation facilitates the partial inclusion of turbulence-soot interactions by considering the impact of \dmm{sub-filter} fluctuations in the mixture fraction on the soot source terms. To prevent non-physical soot consumption when $\widetilde{Y}_{s,j}$ approaches zero, the consumption term in Eq.~\eqref{eqn:omegaYs_FGM-T} is linearized as a function of $\widetilde{Y}_{s,j}$. \ajk{The linear-relaxation formulation used here is motivated by the distinct scaling of the soot subprocesses. Since soot oxidation dominates the overall soot consumption, a linear relaxation model provides a reasonable approximation for the soot consumption rates, as demonstrated in~\cite{kalbhor2023}.}\ajk{ Note that the various subprocesses involved in soot formation depend not only on local gas-phase conditions but also on the evolving soot characteristics (e.g. soot number density). In the FGM-T formulation, the non-linear dependence of the soot source terms on the soot variables is fully accounted for within the flamelet calculations rather than in the CFD calculation. The FGM-T model therefore does have limitations in capturing the full chemical trajectories associated with soot formation, particularly the transition from gas-phase precursors to steady-state soot, as these pathways are not explicitly retained in the manifold.} A detailed description of this model and its underlying approximations can be found in Kalbhor et al.~\cite{kalbhor2023}. More details on both the FGM-C and FGM-T approaches, in the context of LES, can be found in Kalbhor et al.~\cite{kalbhor2024}.

\section{Experimental test rig}
\label{sec:case_study}

This study focuses on a bluff body, swirl-stabilized, non-premixed ethylene flame that has been experimentally characterized at the University of Cambridge. The following sections describe the gas turbine combustor model and the numerical setup used for the simulations.

\subsection{Gas turbine combustor model}

The flame is investigated in a lab-scale gas turbine combustor model operating at atmospheric pressure \cite{elhelou2021,defalco2021}. The combustor has a square cross-section of 97\! $\times$\! \SI{97}{mm^2} and a length of \SI{150}{mm}. Air at ambient temperature flows through an axial swirler and enters the combustion chamber through an annular passage formed by a \SI{25}{mm} diameter bluff body and a \SI{37}{mm} diameter outer pipe wall. Gaseous ethylene, also at ambient temperature, is injected through a central fuel inlet with a diameter of \SI{4}{mm}. \lppr{The condition examined corresponds to the baseline case without secondary air dilution. The fuel inlet velocity ($U_f$) is \SI{15.0}{m/s}, and the bulk air inlet velocity ($U_a$) is \SI{15.8}{m/s}, resulting in a global equivalence ratio ($\phi_g$) of 0.3.}

The validation of the gas-phase results relies on line-of-sight (LOS) \ce{OH^*} chemiluminescence and \ce{OH} planar laser-induced fluorescence (PLIF) measurements provided by El Helou et al. \cite{elhelou2021}. The validation of the soot predictions is based on laser-induced incandescence (LII) measurements of the soot volume fraction ($fv$), \lpp{extinction measurements}, and \textit{in situ} probe PSD measurements from El Helou et al. \cite{elhelou2021} and De Falco et al. \cite{defalco2021}.

\subsection{Numerical setup}

The high-fidelity LES simulations are performed using Alya~\cite{vazquez2016}, a multi-physics finite element solver developed at the Barcelona Supercomputing Center (BSC). Alya employs a low-dissipation conservative finite element scheme for low-Mach reacting flows~\cite{both2020}. Stabilization of the continuity equation is achieved through a non-incremental fractional-step approach, modified for variable-density flows. Time integration is carried out using a third-order Runge-Kutta scheme for momentum and scalar transport. A standard convective outflow condition is applied at the outlet, while a steady inflow boundary condition is imposed at the air inlet of the combustor. At the fuel inlet, synthetic turbulence is introduced following the approach described by Kempf et al. \cite{kempf2005}. Heat losses are accounted for through two mechanisms. First, gas-phase radiation is modeled using a radiative heat loss source term incorporated into the energy transport equation based on the optically thin approximation~\cite{govert2017}. Second, convective heat losses to the walls are introduced by imposing isothermal wall boundary conditions.

\begin{figure}
\centering\includegraphics[width=0.6\linewidth]{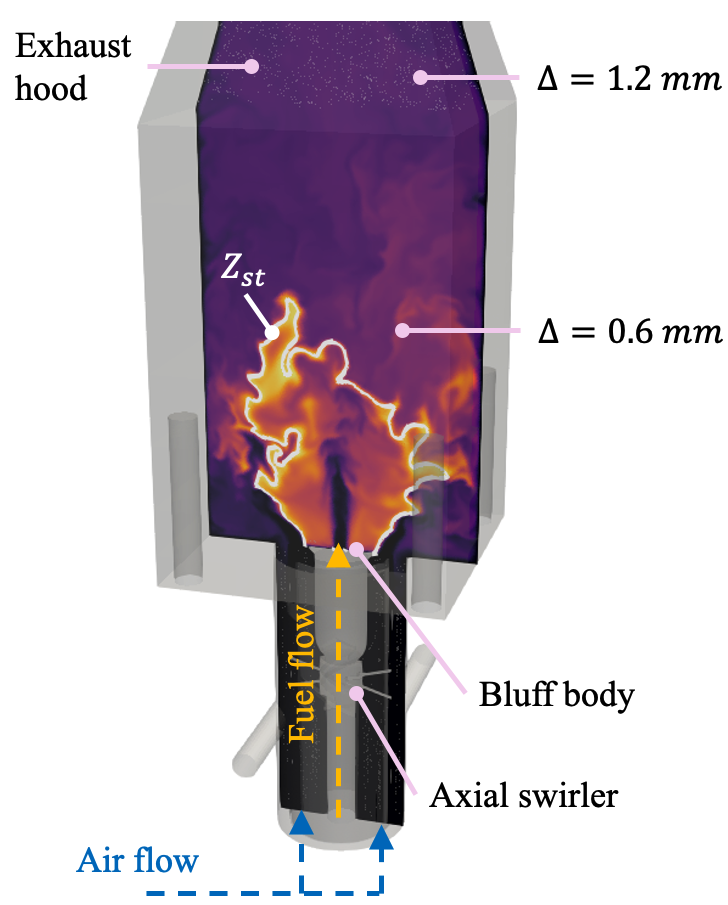}
\caption{Computational domain including a slice placed at the center of the combustor colored by instantaneous temperature. $Z_{st}$ isocontour in white.}\label{fig:mesh}
\end{figure}

The computational domain is discretized using a hybrid mesh with approximately 10 million degrees of freedom. As shown in Fig.~\ref{fig:mesh}, the domain includes key components such as the air and fuel inlets, axial swirler, bluff body, combustion chamber, and exhaust hood. Four pipes, located at the corners of the combustion chamber, are designed for the introduction of dilution air, although this feature is not considered in the present study. The minimum element size is \SI{0.6}{mm} in the combustion chamber and \SI{1.2}{mm} in the rest of the domain, as illustrated in a mesh slice colored by the instantaneous temperature field. A white line, drawn along the isocontour of $Z_{st}$, highlights the stoichiometric mixture fraction. The mean results presented in the following sections are first time-averaged and subsequently azimuthally averaged with a resolution of one degree. Statistical data is collected over three flow-through times. \lppr{The computational cost of simulating \SI{10}{ms} of physical time is in the order of $65\times10^3$ CPU hours for FGM-C and $11\times10^3$ CPU hours for FGM-T.}

\clearpage

\section{Results and discussion}
\label{sec:results}

Soot predictions are analyzed with a focus on key aspects addressed throughout the next four sections with a comprehensive comparison with experimental data~\cite{elhelou2021,defalco2021}. Section~\ref{subsec:soot_structure}, addressing the spatial structure of soot volume fraction across the combustor. \lppr{Section~\ref{subsec:soot_psd}, with the analysis of the predicted PSD and Section~\ref{subsec:flow_soot}, describing the interaction between the flow field and the soot phase.} \dmm{Finally, Section~\ref{subsec:soot_intermittency} introduces} the characterization and quantification of soot intermittency. Gas-phase predictions and validation are provided as part of the supplementary materials.

\subsection{Soot spatial structure}
\label{subsec:soot_structure}

\begin{figure}
\centering\includegraphics[width=0.7\linewidth]{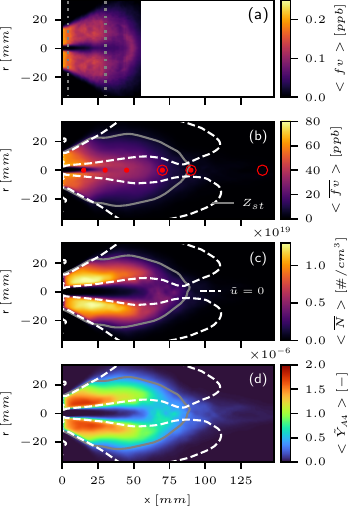}
\caption{\lppr{Experimental LII soot volume fraction~\cite{defalco2021} (a). FGM-C predictions of mean soot volume fraction (b), soot number density (c), and \ce{A}4 mass fraction (d). The stoichiometric mixture-fraction isocontour is shown in gray, and the zero-axial-velocity isocontour in white. Red points and circles indicate the locations of extinction and \textit{in-situ} PSD measurements, respectively.}}\label{fig:field_fv}
\end{figure}

\begin{figure}
\centering\includegraphics[width=0.5\linewidth]{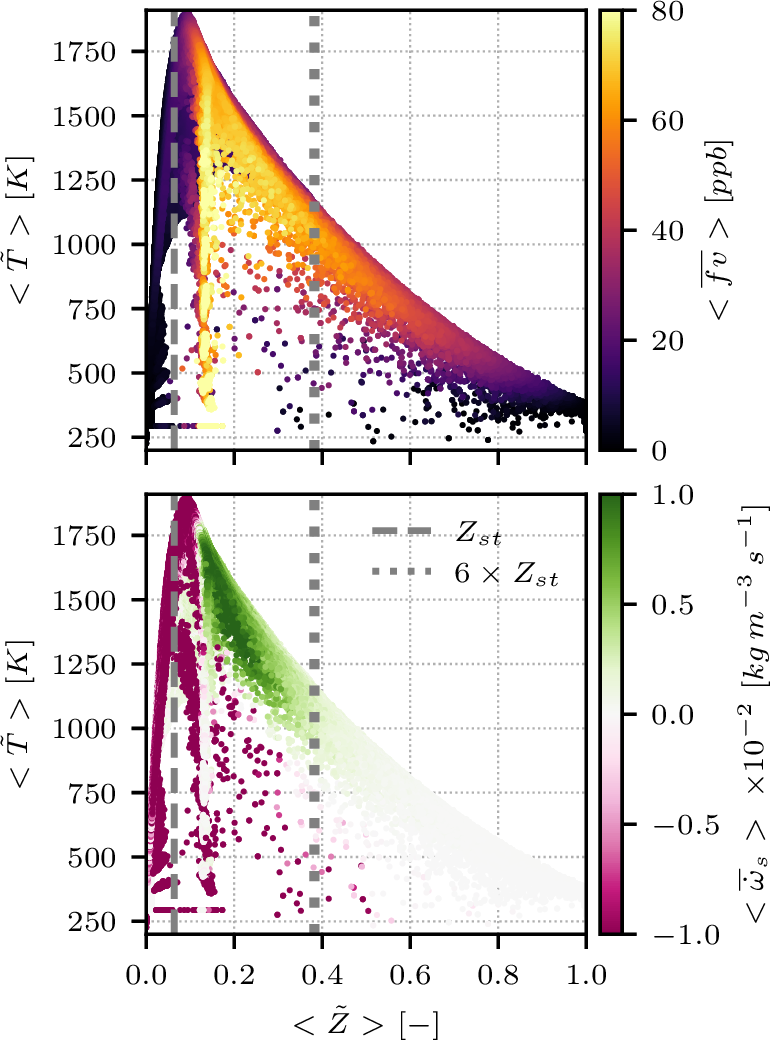}
\caption{\lppr{Predicted mean mixture-fraction–temperature scatter colored by mean soot volume fraction (top) and soot mass-fraction source term (bottom). Gray vertical lines indicate $Z_{st}$ (dashed) and $6 \times Z_{st}$ (dotted). Results correspond to the FGM-C approach.}}\label{fig:scatter_fv}
\end{figure}

\begin{figure}
\centering\includegraphics[width=0.4\linewidth]{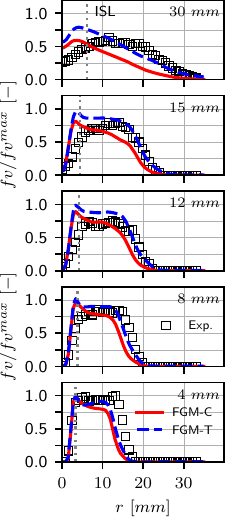}
\caption{\lppr{Predicted mean normalized soot volume fraction radial profiles from FGM-C (red) and FGM-T (blue) approaches. A dotted gray line indicates the location of the ISL. Normalized experimental LII data from~\cite{defalco2021}.}}\label{fig:fv_profiles}
\end{figure}

\begin{figure}
\centering\includegraphics[width=0.5\linewidth]{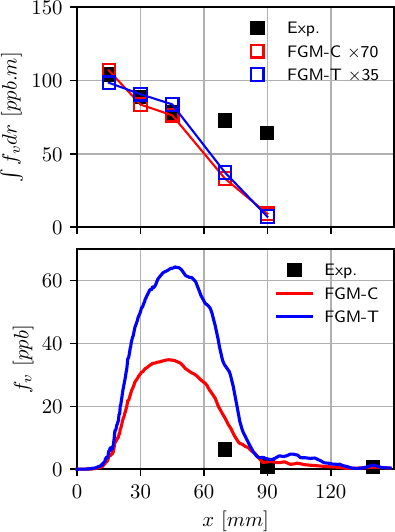}
\caption{\lppr{Mean results along the combustor axis. Predicted LOS soot volume fraction and extinction measurements (top). Predicted soot volume fraction compared with experimental values derived from PSD measurements (bottom). FGM-C predictions are shown in red and FGM-T in blue. Experimental data from~\cite{defalco2021}.}}\label{fig:fv_axis}
\end{figure}

\lppr{The spatial distribution of soot across the combustor is shown in Fig.~\ref{fig:field_fv} for both experiments and simulations. Panel (a) presents the mean soot volume fraction obtained from LII measurements~\cite{defalco2021}, while panels (b), (c), and (d) provide the corresponding LES predictions of mean $\overline{fv}$, soot number density, and \ce{A}4 mass fraction. In Fig.~\ref{fig:field_fv} (a), the gray dashed lines indicate the radial locations used for the profiles in Fig.~\ref{fig:fv_profiles}. In Fig.~\ref{fig:field_fv} (b), the red points and circles denote the positions of the extinction and \textit{in-situ} PSD measurements shown in Fig.~\ref{fig:fv_axis} and Fig.~\ref{fig:PSD}, respectively. Overall, the LES reproduces the experimental soot distribution, capturing the highest soot levels near the bluff body at the flame base and the gradual decay toward the burner exit. This agreement is consistent with previous FGM-T simulations, which reported similar spatial soot trends~\cite{pachano2024}. In Fig.~\ref{fig:field_fv} (b), the predicted $\overline{fv}$ field is confined to the fuel-rich region, bounded by the stoichiometric mixture-fraction isocontour (solid gray line). The zero-axial-velocity isocontour (dashed white line) further indicates that the primary sooting region lies within the recirculation zone. Within this zone, Fig.~\ref{fig:field_fv} (c) shows an elongated region of peak number density near the inner shear layer (ISL) and around $x\approx$\SI{25}{mm}. This location coincides with the maximum \ce{A}4 concentration in Fig.~\ref{fig:field_fv} (d), highlighting the dominant contribution of nucleation to the number-density peak under these conditions.}

\lppr{Whereas Fig.~\ref{fig:field_fv} describes the spatial distribution, Fig.~\ref{fig:scatter_fv} presents the distribution in composition space. The top panel shows the scatter of time-averaged mixture fraction and temperature colored by $\overline{fv}$, while the bottom panel shows the same scatter colored by $\overline{\dot{\omega}}{s}$. Figure~\ref{fig:scatter_fv} indicates that $\overline{\dot{\omega}}{s}$ attains its highest values within the fuel-rich region for $\tilde{Z} < 6 Z_{st}$ under intermediate- to high-temperature conditions ($\tilde{T} > \SI{1000}{K}$). These conditions correspond to locations where soot concentrations are highest. At the same time, Fig.~\ref{fig:scatter_fv} also shows that large amounts of soot extend across a wider temperature range, reaching values close to those near the cold wall at the burner base, consistent with the spatial distribution in Fig.~\ref{fig:field_fv}.}

\begin{figure}
\centering\includegraphics[width=0.7\linewidth]{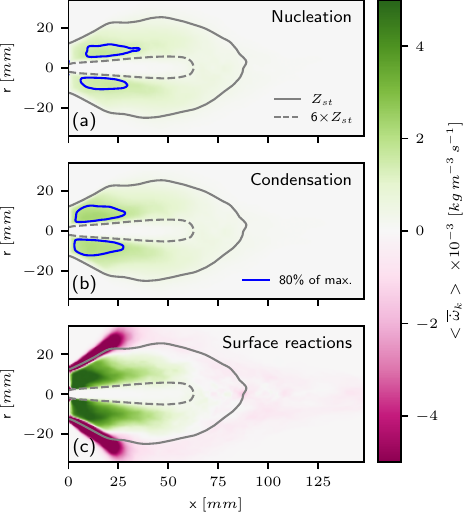}
\caption{\lppr{Predicted mean soot source terms from nucleation (a), condensation (b), and surface reactions (c). Gray lines indicate $Z_{st}$ (solid) and $6 \times Z_{st}$ (dashed) isocontours. Blue isocontour drawn at 80\% of peak value. Results correspond to the FGM-C approach.}}\label{fig:field_sources}
\end{figure}

\lppr{To further characterize the spatial distribution of soot, the time-averaged source terms associated with the different soot formation processes are presented in Fig.~\ref{fig:field_sources}. These include nucleation (a), condensation (b), and surface reactions encompassing both growth and oxidation (c). Although coagulation is also part of the soot sub-processes represented in the model, it is not included in the global soot mass balance because it does not change the total soot mass fraction. Coagulation is strictly mass-conservative as it redistributes soot among particle-size sections, modifying the PSD and, indirectly, the surface area available for surface-reaction processes. Its influence on the mass-based formulation therefore enters only through this second-order effect.}

Source terms directly contributing to soot production namely nucleation, condensation, and the positive component of the surface reaction source term are predominantly confined within the fuel-rich region, as delineated by the stoichiometric isocontour (solid gray line). Notably, within this highly fuel-rich region ($6\times Z_{st}$ isocontour depicted with a dashed gray line) , soot formation is suppressed. To highlight the spatial extent of nucleation and condensation, 80\% peak-value isocontours are displayed with blue lines in Fig.~\ref{fig:field_sources} (a) and (b), respectively. As anticipated, these processes are most intense at locations corresponding to peak \ce{A}4 values \lppr{(see Fig.~\ref{fig:field_fv} (d))}. Interestingly, the nucleation peak is spatially offset from the condensation peak, suggesting nucleated soot particles grow through condensation and surface reactions as they are convected. In this specific case, the flow is channeled towards the bluff body by recirculation, thus promoting significant soot growth in the adjacent region. Oxidation, represented by the negative component of the surface reaction source term in Fig.~\ref{fig:field_sources} (c), primarily occurs near the stoichiometric isocontour. This process is particularly significant in the direction of the incoming swirling air (see gas-phase results in supplementary materials). Quantitatively, oxidation exhibits the strongest influence, with a peak value of \lppr{$-2.12\times10^{-1}$ \SI{}{kg.m^{-3}.s^{-1}}}. Please note that the color range in Fig.~\ref{fig:field_sources} is limited to \lppr{$\pm5.0\times10^{-3}$ \SI{}{kg.m^{-3}.s^{-1}}} to facilitate the comparison between source terms. Among the production processes, surface reactions driven by HACA dominate soot growth, reaching a peak value of \lppr{$9.48\times10^{-3}$ \SI{}{kg.m^{-3}.s^{-1}}}, followed by condensation \lppr{($2.26\times10^{-3}$ \SI{}{kg.m^{-3}.s^{-1}})} and nucleation \lppr{($1.79\times10^{-3}$ \SI{}{kg.m^{-3}.s^{-1}})}. This hierarchy of soot subprocesses, based on peak values, aligns with previous observations indicating surface reactions are the primary driver of soot production in recirculation-dominated flows~\cite{chong2018}, in contrast with turbulent jet flames where condensation is the dominant process~\cite{ferraro2022}.

\lppr{The spatial distribution of soot is further assessed through the mean soot volume fraction profiles shown in Fig.~\ref{fig:fv_profiles}, where results from the two modeling approaches, FGM-C and FGM-T, are compared against the experimental data. Profiles are extracted at six axial locations between 4 and \SI{30}{mm}, indicated by the gray dashed lines in Fig.~\ref{fig:field_fv}(a). Consistent with the strong spatial agreement observed in Fig.~\ref{fig:field_fv} for the FGM-C predictions, the radial profiles reproduce the overall shape of the experimental measurement, also in line with previous FGM-T results reported in~\cite{pachano2024}. To enable a direct comparison across all datasets, each profile is normalized by the peak value at \SI{4}{mm}, where the soot concentration is highest. The corresponding peak values are \SI{0.1690}{ppb} for the experiment, \SI{65.9}{ppb} for FGM-C, and \SI{92.7}{ppb} for FGM-T. When expressed in this normalized form, the profiles in Fig.~\ref{fig:fv_profiles} clearly show the progressive decay of soot volume fraction with increasing axial distance, a trend captured by both modeling approaches.

At \SI{4}{mm}, the peak soot volume fraction occurs near the ISL, marked by the dotted gray line. As shown in Fig.~\ref{fig:field_fv}, this region exhibits elevated \ce{A}4 concentrations, which promote nucleation and result in the observed peak in soot number density. The source terms in Fig.~\ref{fig:field_sources} further indicate that condensation and surface growth are also strong in the vicinity of the ISL, explaining the sharp increase in the \SI{4}{mm} profile around $r \approx \SI{3}{mm}$. Further downstream, the predicted peak remains located near the ISL, while the experimental peak shifts slightly outward, a difference most apparent at \SI{30}{mm}. This may suggest that the radial expansion of the ISL occurs more slowly in the LES than in the experiment. These observations apply to both modeling approaches, confirming that they produce comparable soot volume fraction spatial structures in line with the main experimental trends.}

The integrated \lpp{$\overline{fv}$} from the LES, scaled to facilitate comparison with the experimental data, is shown in Fig.~\ref{fig:fv_axis} (top). Between 15 and \SI{45}{mm}, the axial decay of soot predicted by the simulations agrees well with the experimental measurements. \lppr{Further downstream, however, the numerical results exhibit a stronger reduction in soot volume fraction, which may be linked to the dominant influence of soot oxidation—primarily through surface reactions with \ce{OH} radicals—particularly in the stoichiometric region (around \SI{90}{mm}) where \ce{OH} concentrations are higher. A similar tendency for soot oxidation to progress more rapidly in the simulations than in the experiments was also reported for the FGM-T formulation in~\cite{pachano2024}. Since this behavior is likewise observed with FGM-C, it likely reflects limitations in the soot model rather than in the LES coupling strategy. A more detailed soot representation, for instance through a bi-variate formulation, may help assess the impact of better resolving soot surface–related processes, including oxidation, in this configuration.}

Similarly, the predicted on-axis \lpp{$\overline{fv}$} profile in Fig.~\ref{fig:fv_axis} (bottom) shows a pronounced decrease beyond \SI{60}{mm}, indicating that soot is predominantly oxidized at the \lpp{locations of the} \textit{in-situ} PSD measurements. At \SI{70}{mm}, the measured $fv$ is 6.5 $\pm$ \SI{1.0}{ppb}, whereas the FGM-C and FGM-T predictions yield 16.28 and \SI{32.95}{ppb}, respectively. Beyond this point, experimental values drop to the order of \lpp{$\mathcal{O}(10^0)$} \SI{}{ppb}, while LES results decrease to the order of \lpp{$\mathcal{O}(10^{-2})$} to \lpp{$\mathcal{O}(10^{-3})$} \SI{}{ppb}. Furthermore, Fig.~\ref{fig:fv_axis} illustrates that FGM-C predicts lower soot concentrations than FGM-T, consistent with differences previously observed for a turbulent non-premixed jet flame \cite{kalbhor2024}. 
\ajk{In the FGM-C approach, soot sub-processes and their rates are explicitly computed using the local soot variables (e.g. mass fraction, number density), providing a more realistic qualitative description of unsteady soot evolution in LES. In contrast, FGM-T soot source terms are calculated and stored for soot mass fractions in a steady-state flamelet. As a result, trajectories concerning the formation of soot from the gas-phase to the steady state are not explicitly retained in the FGM-T strategy. Furthermore, the nonlinear dependence of soot production rates on soot variables is not accounted for in the current FGM-T formulation. Consequently, directly retrieving soot production rates from the table may lead to 
higher values compared to the FGM-C approach.}
\lppr{Quantitative validation of soot volume fraction in lab-scale gas turbine combustor models remains challenging due to the inherent complexities of experimental measurements. As shown by the dataset used for validation \cite{defalco2021}, different diagnostics capture complementary but not identical portions of the soot population. For instance, LII-derived soot volume fractions may differ considerably from values obtained through in-situ probe sampling, reflecting the higher sensitivity of LII to larger particles and the broader size range accessed by probe-based techniques. In line with the discussion in \cite{pachano2024}, the analysis therefore focuses on spatial trends and relative variations rather than absolute magnitudes. The agreement in these trends between LES, both FGM-C and FGM-T, and the experiments indicates that the model captures the dominant physical processes governing soot formation and its spatial distribution.}

\clearpage

\subsection{Analysis of particle size distributions}
\label{subsec:soot_psd}

\begin{figure}
\centering\includegraphics[width=0.5\linewidth]{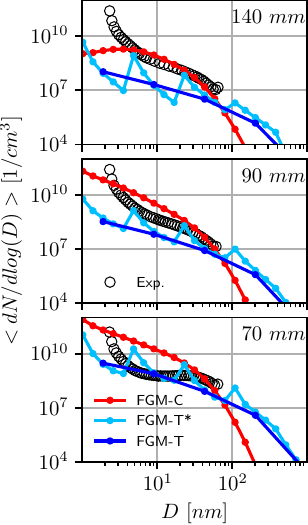}
\caption{\lppr{Mean PSD results along the combustor axis. Predictions from FGM-C (red) and FGM-T (blue), compared with measurements from~\cite{defalco2021}.}}\label{fig:PSD}
\end{figure}

In addition to soot volume fraction values shown in Fig.~\ref{fig:fv_axis} (bottom), the \textit{in situ} measurements from \cite{defalco2021} provide valuable insights into PSD evolution, which is rarely accessible in turbulent flames from lab-scale gas turbine combustors. 
\dmm{Time-averaged PSD along the axis at three locations (see red circle markers in Fig.~\ref{fig:field_fv} (b)) are shown in Fig.~\ref{fig:PSD} for simulations and experiments. Results from both FGM-C and FGM-T are included for comparison. Note that while the FGM-C approach directly provides the PSD (depicted in red), the PSD obtained with the FGM-T approach corresponds only to the clustered soot sections (depicted in blue).}
The results labeled as \say{FGM-T*} represent the original thirty soot sections \dmm{distribution} from FGM-T, reconstructed in a post-processing step. \abk{The observed discontinuities at cluster boundaries are attributed to limitations of the soot PSD reconstruction method in low-soot concentration regions and the assumption made for soot section clustering, as presented in the original formulation~\cite{kalbhor2023}.} Across all three measurement locations, both experimental and numerical PSD exhibit a unimodal distribution, with LES predictions for both modeling approaches showing reasonable agreement in terms of particle count. 

\begin{figure}
\centering\includegraphics[width=0.7\linewidth]{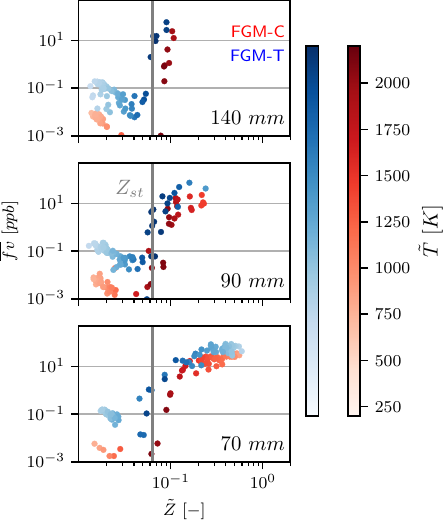}
\caption{ \lppr{Predicted mixture-fraction-soot-volume-fraction scatters colored by temperature. FGM-C results in red and FGM-T in blue. Vertical gray line marks $Z_{st}$.}}\label{fig:fv_scatter_PSD}
\end{figure}

Further insights into the PSD at these locations can be gained from the mixture state shown in Fig.~\ref{fig:fv_scatter_PSD}, where the $\tilde{Z}$-$\overline{fv}$ scatter is colored by temperature. These results indicate that while the mixture is, on average, fuel-rich at \SI{70}{mm} and lean at \SI{90}{mm} and \SI{140}{mm}, both lean and rich \dmm{reacting fronts} are observed at each location. Additionally, they reveal that most soot at the latter two axial positions originates from intermittent events where the thermochemical state corresponds to fuel-rich, high-temperature conditions.

\clearpage

\subsection{Flow-soot interactions}
\label{subsec:flow_soot} 

\begin{figure}
\centering\includegraphics[width=0.85\linewidth]{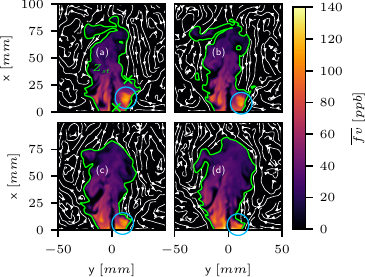}
\caption{\lppr{Instantaneous soot volume fraction fields (a–d) predicted using the FGM-C approach. The $Z_{st}$ isocontour is shown in green, and white streamlines are derived from the velocity field.}}\label{fig:field_inst_fv}
\end{figure}

\lppr{While peak soot source terms correlate with the spatial distribution of $\overline{fv}$, understanding the underlying mechanisms requires a more detailed examination of the flow field. Figure~\ref{fig:field_inst_fv} shows the instantaneous evolution of the velocity field, illustrated through white streamlines, and the corresponding $\overline{fv}$ distribution, with an effective temporal resolution of approximately \SI{2}{ms}. A close inspection of these snapshots highlights the central role of the recirculation vortex in shaping the soot field. In particular, the region marked by a blue circle contains soot-rich gases that are continuously advected toward the burner base by the recirculation vortex. This soot pocket is subsequently compressed into a thin layer along the wall, as illustrated in Fig.~\ref{fig:field_inst_fv} (d).}

\begin{figure}
\centering\includegraphics[width=0.6\linewidth]{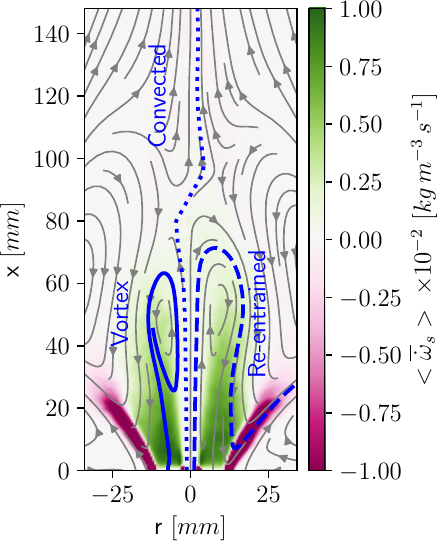}
\caption{\lppr{Predicted mean net soot mass fraction source term using the FGM-C approach. Stream lines derived from the mean velocity field depicted in gray. Blue lines show the trajectory of key streamlines \textit{i.e.}, re-entrained (dashed), convected (dotted), and vortex (solid).}}\label{fig:field_net_omegaYs}
\end{figure}

\begin{figure}
\centering\includegraphics[width=0.6\linewidth]{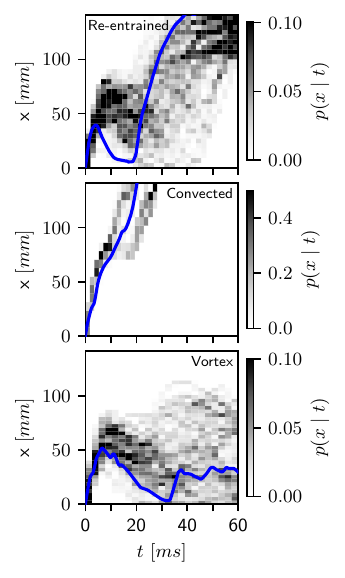}
\caption{\lppr{Time-conditioned PDFs of the axial coordinate for the three trajectory classes: re-entrained (top), convected (middle), and vortex (bottom). In each panel, the trajectory with the highest peak $\overline{fv}$ within the corresponding class is highlighted in blue. Results correspond to the FGM-C approach.}}\label{fig:lagrangian_trajectories}
\end{figure}

\lppr{While these instantaneous fields already indicate a strong link between the flow structures and the resulting soot distribution, Fig.~\ref{fig:field_net_omegaYs} provides further insight by superimposing streamlines on the mean $\overline{\dot{\omega}}_{s}$ field. Three characteristic streamline families that govern the soot distribution can be identified. Two of them interact directly with the recirculation vortex. The first, denoted \say{Re-entrained} (dashed blue), corresponds to fluid elements released near the fuel inlet. These elements are initially transported toward the outlet but are later recirculated back toward the burner base before following the swirling motion and exiting around $y\approx$ \SI{50}{mm}. The second family, denoted \say{Vortex} (solid blue), represents fluid elements released from the recirculation vortex that are transported directly toward the burner base. A third family, \say{Convected} (dotted blue), consists of fluid elements that are continuously transported toward the outlet without significant interaction with the recirculation region.}

\lppr{These streamline patterns provide an intuitive interpretation of the global flow–soot interactions and form a basis for understanding the spatial structure of $\overline{fv}$. However, analysis based solely on mean streamlines is limited, as the actual trajectory of a fluid element in a turbulent flow often deviates from the mean path. Consequently, previous studies on soot production in turbulent flows point to Lagrangian trajectories as a more suitable framework. The validity of this approach has been assessed in DNS studies of non-premixed jet flames~\cite{Attili.2013,attili2014}, where soot was tracked directly in the Lagrangian framework. Later, the suitability of particle-like trajectories, based on tracking notional particles advanced via local fluid velocity, was demonstrated in LES studies of the DLR combustor \cite{chong2018}.}

\lppr{To analyze the history of fluid elements representative of the turbulent flow evolution, massless particles are tracked following the framework in \cite{chong2018}. These particle-like trajectories result from tracking particles seeded from concentric circles with radii of 0.5, 1.0, 1.5, and \SI{2.0}{mm}. These circles are located in the $zy$-plane just downstream of the combustor backplane at $x=\SI{0.5}{mm}$. Particles are released with a ten-degrees angular resolution and tracked for \SI{60}{ms}.} \lppr{The resulting Lagrangian trajectories are then classified according to their large-scale axial evolution, consistent with the streamline-based topology in Fig.~\ref{fig:field_net_omegaYs}. Trajectories ending upstream of \SI{100}{mm} are classified as vortex type. Trajectories ending downstream are assigned to either re-entrained or convected. A trajectory is considered re-entrained if its minimum axial position during the first \SI{10}{ms} falls below \SI{70}{mm}, otherwise it is classified as convected. Figure~\ref{fig:lagrangian_trajectories} shows the time-conditioned PDF of the axial coordinate for each class, with the trajectory exhibiting the highest peak $\overline{fv}$ in each class highlighted in blue.}

\begin{figure}
\centering\includegraphics[width=0.95\linewidth]{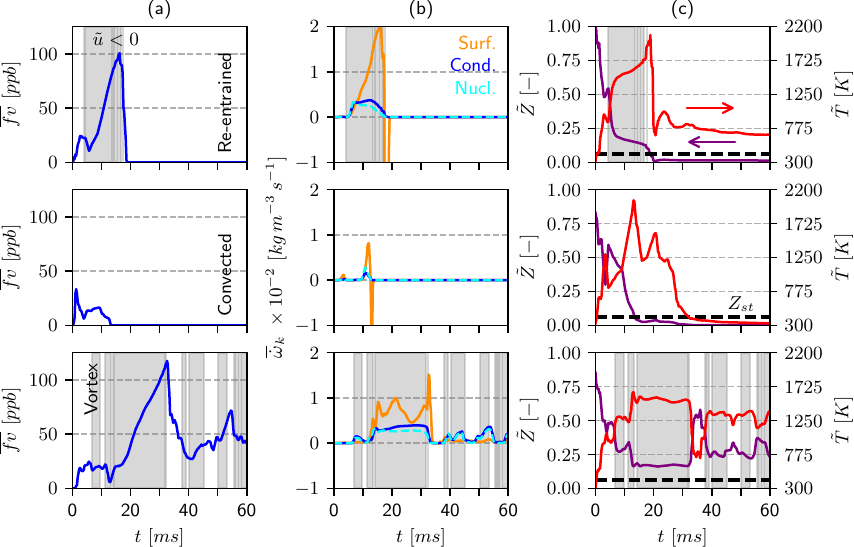}
\caption{\lppr{Predicted instantaneous variables along representative trajectories. (a) Soot volume fraction. (b) Source terms for surface reactions (orange), condensation (blue), and nucleation (cyan). (c) Mixture fraction (purple, left axis) and temperature (red, right axis). The gray shaded region indicates the period during which the trajectory remains within the recirculation zone. Results correspond to the FGM-C approach.}}\label{fig:time_streamlines}
\end{figure}

\lppr{To analyze these representative trajectories, Fig.~\ref{fig:time_streamlines} presents the temporal evolution of $\overline{fv}$ and the variables governing its production. Panel (a) shows $\overline{fv}$, panel (b) displays source terms (surface reactions, condensation and nucleation), and panel (c) shows mixture fraction (left axis) and temperature (right axis). Time intervals during which the trajectory resides in the recirculation zone ($\tilde{u}<0$) are shaded in gray.}

\lppr{Initially, all three trajectory types experience a similar evolution as the tracking of massless elements begins within the fuel jet or its immediate vicinity. Consequently, within the first \SI{5}{ms}, the trajectories exhibit an increase in $\overline{fv}$ as the fluid element is advected downstream. It undergoes progressive dilution with hot combustion gases, transitioning from cold $\tilde{Z}\approx1$ conditions towards hotter, leaner conditions containing soot transported from more productive regions. Among the three, the convected trajectory exhibits the lowest $\overline{fv}$ because it transits rapidly through the region favorable for soot production. Around $t\approx10$ \SI{}{ms}, this trajectory experiences brief pulses of nucleation, condensation, and surface reactions, but the mixture fraction quickly decreases toward $Z_{st}$ as the fluid approaches the flame front, accompanied by a sharp temperature increase above \SI{1800}{K}. Under these conditions soot is fully oxidized, consistent with the strong negative peak in the surface reaction source term and the drop of $\overline{fv}$ to zero.} \lppr{In contrast, the re-entrained and vortex trajectories exhibit a distinct evolution due to their interaction with the recirculation vortex and the consequent transport toward the burner base. There, the mixture fraction remains approximately constant ($\tilde{Z}\approx0.16$), and the temperature stabilizes at around $\tilde{T}\approx1650$. These conditions are highly favorable for soot production, as reflected in the source-term profiles: nucleation and condensation persist with comparable magnitudes for both trajectories, and soot grows predominantly through surface reactions, consistent with the mean fields in Fig.~\ref{fig:field_net_omegaYs}.}

\lppr{A later stage distinguishes the two trajectories. In the re-entrained case, the fluid element eventually crosses the flame front, causing $\tilde{Z}$ to fall below $Z_{st}$. This leads to a temperature rise up to \SI{2082}{K} and strong oxidation, visible as a sharp negative surface-reaction peak. The vortex trajectory, however, remains on the fuel-rich side of the flame ($\tilde{Z} > Z_{st}$) preventing oxidation. The decline in $\overline{fv}$ at the end of the first recirculation period arises instead from mixing with colder, richer gases, indicated by the sharp rise in $\tilde{Z}$(up to $>$ 0.5) and the corresponding temperature drop to $\tilde{T}\approx775$ \SI{}{K}. Beyond this point, the trajectory undergoes intermittent re-entrainment events, each contributing to incremental increases in $\overline{fv}$. During these later cycles, the mixture fraction remains slightly richer ($\tilde{Z}\approx0.25$), and the temperatures slightly lower, resulting in conditions less favorable for sustained surface-reaction growth. Consequently, the source terms display intermittent pulses in which nucleation and condensation periodically exceed surface reactions.}

\lppr{The instantaneous evolution along the Lagrangian trajectories demonstrates that soot production is primarily governed by the thermochemical conditions encountered within the recirculation vortex. The sustained growth, heavily favored by surface reactions, and the subsequent accumulation of soot as fluid elements are transported toward the burner base, explains the presence of the highest soot concentrations at that location, as observed in Fig.~\ref{fig:field_fv} in both simulations and experiments.}
  
\subsection{Intermittency}
\label{subsec:soot_intermittency}

\lpp{Instantaneous flow-soot interactions depicted in Fig.~\ref{fig:field_inst_fv} visually confirm the intermittent nature of soot production, as also suggested by \lpp{$\overline{fv}$} fluctuations in Fig.~\ref{fig:fv_scatter_PSD}.} The green \say{X} markers in Fig.~\ref{fig:field_inst_fv} help guide the reader through this intermittency. At the first marked location ($x = \SI{30}{mm}$, \lpp{$y = \SI{15}{mm}$}), the mixture is initially close to stoichiometry, as indicated by its position in the vicinity of the $Z_{st}$ isocontour. Over time, the mixture transitions to a lean state with no soot as the recirculation vortex pushes the stoichiometric isosurface (Fig.~\ref{fig:field_inst_fv} (a–c)). At the second marked location ($x = \SI{4}{mm}$, \lpp{$y = \SI{4}{mm}$}), the mixture remains fuel-rich at all depicted time instants. However, \lpp{$\overline{fv}$} varies significantly due to soot transport towards the base of the burner. This rapid variation in both mixture fraction and $fv$ over just a few milliseconds is consistently observed throughout the full simulation, and results are provided in the supplementary materials.  

\begin{figure}
\centering\includegraphics[width=0.6\linewidth]{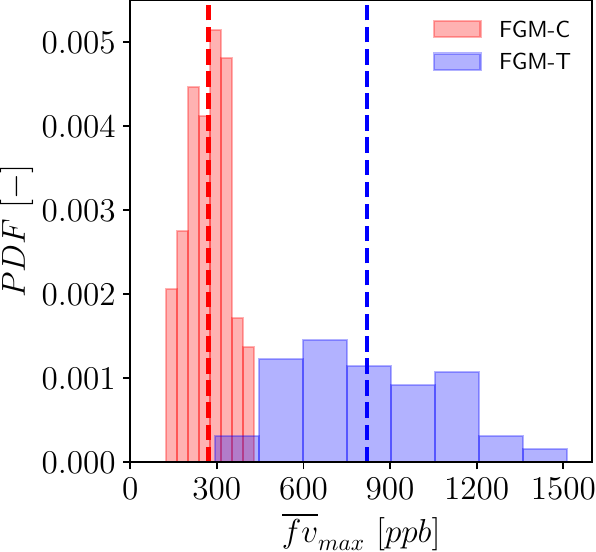}
\caption{\lppr{Predicted marginal PDFs of the instantaneous maximum soot volume fraction. FGM-C results are shown in red and FGM-T in blue. The dashed lines indicate the average value.}}\label{fig:fv-max_PDF}
\end{figure}

\begin{figure}
\centering\includegraphics[width=0.7\linewidth]{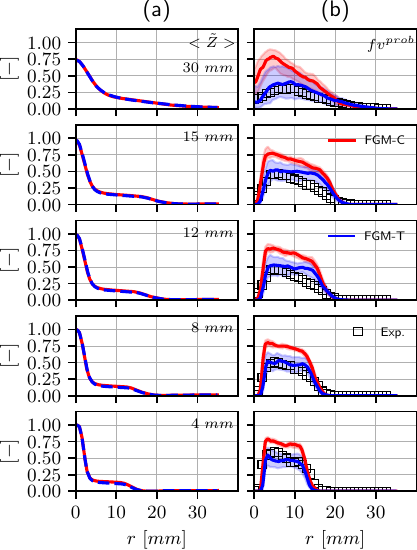}
\caption{\lppr{Predicted (a) mean mixture-fraction radial profiles and (b) mean soot-probability radial profiles. FGM-C results are shown in red and FGM-T in blue. Experimental LII data from~\cite{elhelou2021}.}}\label{fig:fv_prob_profiles}
\end{figure}

Experimental studies have reported soot intermittency for this case \cite{elhelou2021}. In addition to the spatial structure of soot volume fraction, LII measurements provide information on soot intermittency. To quantify this \lpp{phenomena} in the LES, the instantaneous \lpp{$\overline{fv}$} field was binarized (0 or 1) based on a threshold value. Soot probability (\lpp{$\overline{fv}^{prob}$}) was then calculated as the mean of the binarized signal field. Experimentally, the threshold was set to 5\% of the maximum LII signal. For the LES, in order to account for a more representative value, the threshold was determined as the mean value of the instantaneous $\overline{fv}_{max}$, for which the marginal PDF is shown in \lppr{Fig.~\ref{fig:fv-max_PDF}}. \lppr{The mean and standard deviation of the $\overline{fv}_{max}$ time series were 271.3 $\pm$ \SI{72.7}{ppb} for FGM-C and 820.1 $\pm$ \SI{256.7}{ppm} for FGM-T, respectively.} Results for soot intermittency are presented in \lppr{Fig.~\ref{fig:fv_prob_profiles}} at axial locations ranging from 4 to \SI{30}{mm}. Radial profiles of the mean mixture fraction, demonstrating no significant differences between FGM-C and FGM-T are shown in Fig.~\ref{fig:fv_prob_profiles} (a). This is expected, as both modeling approaches rely on the same gas-phase  
\dmm{description,} 
with variations limited to \dmm{only soot calculations}. Soot probability profiles from both LII measurements\footnote{Experimental data at 15 and \SI{30}{mm} were obtained from the work of El Helou et al. \cite{elhelou2021}, while data at 4, 8, and \SI{12}{mm} were provided via private communication with Dr. El Helou and are part of the same dataset.} and LES are depicted in Fig.~\ref{fig:fv_prob_profiles} (b). For the LES, the solid line represents $\overline{fv}^{prob}$ with the threshold set at 5\% of the mean $\overline{fv}_{max}$. The shaded area illustrates the sensitivity of the profile to threshold variations, representing the range defined by 5\% of the mean $\pm$ one standard deviation.

Given the same mixing field (Fig.~\ref{fig:fv_prob_profiles} (a)), both modeling approaches correctly capture soot intermittency, as evidenced by the good correlation between experimental and numerical profiles. These results reinforce the observation that the LES \dmm{with both soot modeling approaches} is able to reproduce the flow and mixing field dynamics, which was also demonstrated by \dmm{the gas-phase analysis conducted in the supplementary materials.} Similar to soot volume fraction profiles in Fig.~\ref{fig:fv_profiles}, both experimental and numerical soot probability profiles show a transition from a top-hat shape at \SI{4}{mm} to a bell-shaped profile at \SI{30}{mm}. Comparing FGM-C and FGM-T results, both models provide reasonable agreement with experimental observations, with greater variability at locations further \dmm{away} from the burner base where mixing and oxidation are more intense. \lppr{As described in Section~\ref{subsec:soot_model}, the FGM-T formulation tabulates soot source terms and integrates them over a presumed subgrid-scale mixture-fraction PDF, providing a statistical treatment of turbulence–soot interaction that is important for capturing intermittency in highly nonlinear soot processes. In contrast, although the FGM-C formulation retains the nonlinear dependence of soot chemistry at runtime, it does not explicitly account for subgrid-scale fluctuations. This difference in subgrid-scale treatment may explain the improved agreement of FGM-T with the measured soot intermittency, particularly at \SI{30}{mm}.} Despite differences in predicting soot probability, both numerical results for $\overline{fv}^{prob}$ and $\overline{fv}$ indicate that soot onset occurs between 15 and \SI{30}{mm} from the bluff body. As shown in Fig.~\ref{fig:field_fv}, the main sooting zone near the bluff body extends radially to approximately \SI{15}{mm} (profile at \SI{4}{mm}), with peak soot values occurring off-axis. Further \lpp{away} from the bluff body, $\overline{fv}^{prob}$ and $\overline{fv}$ profiles widen with increasing axial distance, while the probability of finding soot decreases.

\begin{figure}
\centering\includegraphics[width=0.9\linewidth]{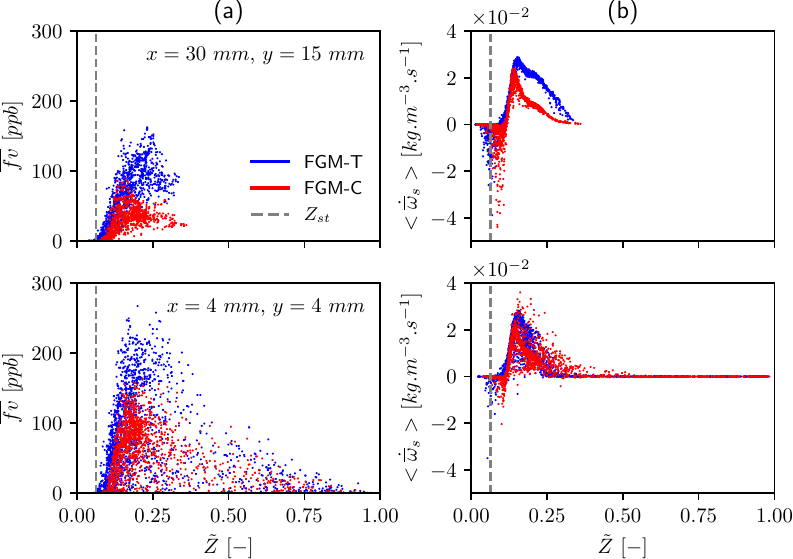}
\caption{\lppr{Predicted scatter of the instantaneous evolution of soot volume fraction (a) and soot mass fraction source term (b) in mixture fraction space for FGM-C (red) and FGM-T (blue). Results are reported at \SI{30}{mm} (top) and \SI{4}{mm} (bottom) from the bluff body.}}\label{fig:intermittency_scatter}
\end{figure}

\begin{figure}
\centering\includegraphics[width=0.9\linewidth]{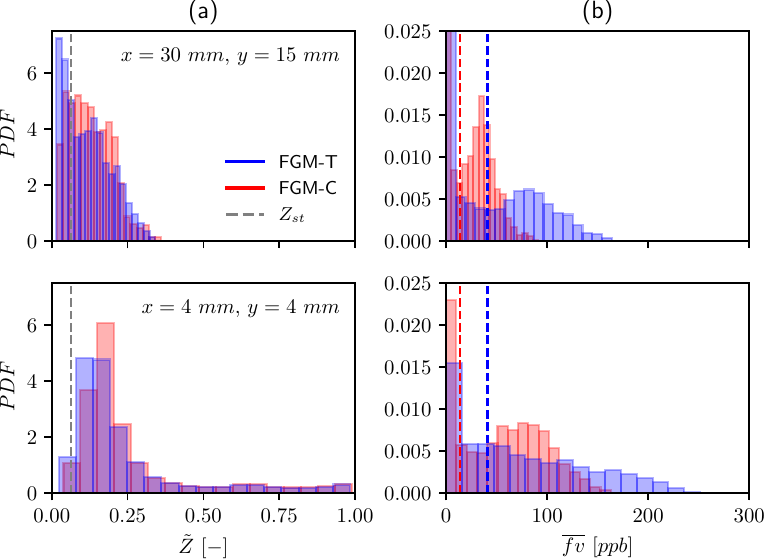}
\caption{\lppr{Predicted marginal PDFs of instantaneous mixture fraction (a) and soot volume fraction (b), evaluated at \SI{30}{mm} (top) and \SI{4}{mm} (bottom) downstream of the bluff body. In (a), the dashed line indicates the stoichiometric mixture fraction, while in (b), the dashed line marks the mean value. FGM-C results are shown in red and FGM-T in blue.}}\label{fig:intermittency_PDF}
\end{figure}

\lpp{A more detailed analysis of the locations denoted by the green markers in Fig.~\ref{fig:field_inst_fv} can yield further insights into soot intermittency.} Both FGM-C and FGM-T predict similar trends in instantaneous \lpp{$\overline{fv}$}, as shown in the scatter plot in Fig.~\ref{fig:intermittency_scatter} (a). These results indicate that while FGM-T predicts a higher overall soot concentration compared to FGM-C, the dispersion in mixture fraction space remains similar between the two modeling approaches. The differences in predicted soot quantities are consistent with variations observed in the scatter plot of \lpp{$\overline{\dot{\omega}}_{s}$} in Fig.~\ref{fig:intermittency_scatter} (b). \abk{This observation is consistent with results for the turbulent non-premixed jet flame studied in \cite{kalbhor2024}.} \dmm{In order to better understand the influence of 
mixture fraction fluctuations on soot intermittency,} Fig.~\ref{fig:intermittency_PDF} presents marginal PDFs for $\tilde{Z}$ (a) and \lpp{$\overline{fv}$} (b). The marginal PDF of $\tilde{Z}$ exhibits \dmm{an expected} unimodal distribution with only minor differences between FGM-C and FGM-T. 
In contrast, the marginal PDF of \lpp{$\overline{fv}$} displays a bimodal distribution, with FGM-T spanning a wider range of \lpp{$\overline{fv}$} values. This broader distribution is consistent with the wider spread observed in the marginal PDF of $\overline{fv}_{max}$ in Fig.~\ref{fig:fv-max_PDF}. Previous studies have identified a correlation between the marginal PDFs of $\tilde{Z}$ and \lpp{$\overline{fv}$} in a turbulent non-premixed jet flame \cite{ferraro2022}. In that case, a bimodal distribution of $\tilde{Z}$ at the jet tip led to a bimodal distribution of \lpp{$\overline{fv}$}. However, the results in Fig.~\ref{fig:intermittency_PDF} suggest that in recirculation-dominated flows, the marginal PDF of $\tilde{Z}$ is decoupled from \lpp{$\overline{fv}$} \dmm{and the intermittency is attributed to the fluctuations of velocity in the recirculation zone.}

\lpp{Previous numerical studies of soot intermittency in lab-scale combustor models, such as the DLR burner, have identified turbulent fluctuations in gas-phase composition as a primary driving mechanism. These fluctuations were shown to be closely linked to the flapping motion of the fuel jet and the resulting detachment of fuel-rich pockets \cite{chong2018,garcia-oliver2024}. Franzelli et al. \cite{franzelli2023} further revealed that flow instabilities, specifically a precessing vortex core (PVC), also contribute to soot intermittency in the DLR burner. Their detailed analysis highlighted the challenges of studying soot intermittency in this configuration due to the low-frequency, long time-scale flow features induced by the PVC. These features necessitated a different approach to validate numerical results, treating them as realizations of possible time-averaged states observed experimentally over equivalent averaging intervals. In contrast to the DLR combustor, the Cambridge combustor exhibits no reported flow instabilities \cite{elhelou2021}. Power spectral density analysis of \ce{OH}-PLIF time-series data within the recirculation zone and along the ISL did not reveal any characteristic frequencies indicative of such instabilities \cite{helou2021}. Experimental findings in this context suggest that the turbulent velocity field influences the intermittent nature of soot, as instantaneous LII measurements show that soot production is not spatially continuous \cite{elhelou2021}. Consistent with this, numerical results presented throughout this work indicate that flow-soot interactions driven by the recirculation vortex affect both the spatial structure and intermittency of soot, \dmm{a phenomenon that both FGM-C and FGM-T modeling approaches were able to capture and } good agreement with the experimental data.}

\section{Summary and conclusions}
\label{sec:conclusions}

\dmm{This study presents a detailed study of soot formation in a swirl-stabilized, non-premixed ethylene flame representative of a lab-scale gas turbine combustor. The methodology proposed is based on the use of LES with a tabulated flamelet model coupled to a discrete sectional soot model. Two modeling strategies for the soot description are considered: i) on-the-fly soot source terms calculation (FGM-C) and ii) fully tabulated \lpp{soot source terms}, clustered-based approach (FGM-T). Description of the approaches are provided and} results are compared with a comprehensive dataset of gas-phase and soot diagnostics, including LII, extinction, and \textit{in situ} PSD measurements.

\dmm{The simulation results showed that both approaches successfully captured the flame structure and key features of the soot distribution, validating the underlying gas-phase chemistry and turbulent combustion framework. Soot formation was found to be highly localized near the bluff body, where low-temperature, fuel-rich conditions and increased residence times, induced by the central recirculation zone, favored soot accumulation despite \abk{relatively modest local soot production rates.}
The analysis revealed that soot dynamics in this configuration are strongly governed by large-scale flow structures, particularly the entrainment and recirculation patterns, which modulate the vortex-induced transport and intermittency of soot. Both FGM-C and FGM-T approaches reproduced the observed intermittent nature of soot, highlighting the role of unsteady vortex–soot interactions. \lppr{Notably, the FGM-T formulation aligns more closely with experimental intermittency data, likely because it better accounts for soot–turbulence interactions by integrating soot source terms over the subgrid-scale $\beta$-PDF of mixture fraction. This result highlights the potential importance of subgrid-scale variability for accurately capturing soot intermittency. A more systematic assessment of the impact of different subgrid-scale closures on soot intermittency is therefore identified as an important direction for future work.} 

Despite their overall consistency, the two models exhibit distinct advantages. The FGM-C approach allows for direct resolution of the soot PSD, which is essential for accurate prediction of radiative properties and particulate mass. However, this comes at a significant computational cost, requiring approximately six times more CPU hours than FGM-T for the same physical time span. Conversely, FGM-T offers a computationally efficient alternative with acceptable accuracy for the mean and fluctuating soot behavior, though it requires assumptions \lpp{for} PSD reconstruction due to soot section clustering.

In conclusion, this work advances the understanding of soot formation in swirl-stabilized combustors by \abk{correlating flow-field characteristics with soot transport and dynamics.} \lppr{It also provides the first comparative assessment between FGM-C and FGM-T with soot clustering of sections in a realistic gas turbine configuration, providing valuable insights for future model developments.}} 

\section*{Declaration of competing interest}

The authors declare that they have no known competing financial interests or personal relationships that could have appeared to influence the work reported in this paper.

\section*{Acknowledgments}

The research leading to these results has received funding from SAFIRE CPP2024-011547, a \say{Proyectos de colaboración público-privada} action of the Spanish Plan Estatal de Investigación Científica, Técnica y de Innovación 2024–2027. Leonardo Pachano acknowledges the AI4S fellowship within the \say{Generación D} initiative, Red.es, Ministerio para la Transformación Digital y de la Función Pública, for talent attraction (C005/24-ED CV1). Funded by the European Union NextGenerationEU funds, through PRTR. DM acknowledges the Grant RYC2021-034654 funded by MICIU\slash AEI\slash 10.13039\slash 501100011033 and by \say{European Union NextGenerationEU/PRTR}. The authors acknowledge computer resources, IM-2023-2-0011 and IM-2023-3-0013, from Red Española de Supercomputación, Spain. \lpp{The authors gratefully acknowledge Dr. Ingrid El Helou and Prof. Epaminondas Mastorakos for generously sharing the experimental data on soot intermittency used in this study.}



\bibliographystyle{elsarticle-num} 
\bibliography{bib}

@article{alzahrani2024,
  title = {International Airport Emissions and Their Impact on Local Air Quality: Chemical Speciation of Ambient Aerosols at {{Madrid}}--{{Barajas Airport}} during the {{AVIATOR}} Campaign},
  shorttitle = {International Airport Emissions and Their Impact on Local Air Quality},
  author = {Alzahrani, Saleh and K{\i}l{\i}{\c c}, Do{\u g}u{\c s}han and Flynn, Michael and Williams, Paul I. and Allan, James},
  year = 2024,
  month = aug,
  journal = {Atmos. Chem. Phys.},
  volume = {24},
  number = {16},
  pages = {9045--9058},
  issn = {1680-7324},
  urldate = {2025-04-30},
  copyright = {https://creativecommons.org/licenses/by/4.0/},
  langid = {english}
}

@article{Appel.2000,
  title = {Kinetic Modeling of Soot Formation with Detailed Chemistry and Physics: {{Laminar}} Premixed Flames of {{C2}} Hydrocarbons},
  author = {Appel, J. and Bockhorn, H. and Frenklach, M.},
  year = 2000,
  journal = {Combust. Flame},
  volume = {121},
  pages = {122--136}
}

@article{Attili.2013,
  title = {Application of a Robust and Efficient {{Lagrangian}} Particle Scheme to Soot Transport in Turbulent Flames},
  author = {Attili, Antonio and Bisetti, Fabrizio},
  year = 2013,
  month = sep,
  journal = {Computers \& Fluids},
  volume = {84},
  pages = {164--175},
  issn = {00457930},
  doi = {10.1016/j.compfluid.2013.05.018},
  urldate = {2025-11-19},
  langid = {english}
}

@article{attili2014,
  title = {Formation, Growth, and Transport of Soot in a Three-Dimensional Turbulent Non-Premixed Jet Flame},
  author = {Attili, Antonio and Bisetti, Fabrizio and Mueller, Michael E. and Pitsch, Heinz},
  year = 2014,
  month = jul,
  journal = {Combust. Flame},
  volume = {161},
  number = {7},
  pages = {1849--1865},
  urldate = {2023-11-23},
  langid = {english}
}

@article{bilger1989,
  title = {The Structure of Turbulent Nonpremixed Flames},
  author = {Bilger, R.W.},
  year = 1989,
  month = jan,
  journal = {Proc. Combust. Inst.},
  volume = {22},
  number = {1},
  pages = {475--488}
}

@article{both2020,
  title = {Low-Dissipation Finite Element Strategy for Low {{Mach}} Number Reacting Flows},
  author = {Both, A. and Lehmkuhl, O. and Mira, D. and Ortega, M.},
  year = 2020,
  month = mar,
  journal = {Comput. Fluids},
  volume = {200},
  pages = {104436},
  urldate = {2022-11-23},
  langid = {english}
}

@article{chong2018,
  title = {Large Eddy Simulation of Pressure and Dilution-Jet Effects on Soot Formation in a Model Aircraft Swirl Combustor},
  author = {Chong, Shao Teng and Hassanaly, Malik and Koo, Heeseok and Mueller, Michael E. and Raman, Venkat and Geigle, Klaus-Peter},
  year = 2018,
  month = jun,
  journal = {Combust. Flame},
  volume = {192},
  pages = {452--472},
  urldate = {2024-04-11},
  langid = {english}
}

@inproceedings{Cokuslu.2022,
  title = {Soot {{Prediction}} in a {{Model Aero-Engine Combustor}} Using a {{Quadrature-based Method}} of {{Moments}}},
  booktitle = {{{AIAA}} 2022-1446},
  author = {Cokuslu, {\"O}mer H. and Hasse, Christian and Geigle, Klaus P. and Ferraro, Federica},
  year = 2022,
  month = jan,
  publisher = {AIAA},
  address = {San Diego, California},
  urldate = {2024-05-06},
  isbn = {978-1-62410-631-6},
  langid = {english}
}

@article{defalco2021,
  title = {Soot Particle Size Distribution Measurements in a Turbulent Ethylene Swirl Flame},
  author = {De Falco, Gianluigi and Helou, Ingrid El and {de Oliveira}, Pedro M. and Sirignano, Mariano and Yuan, Ruoyang and D'Anna, Andrea and Mastorakos, Epaminondas},
  year = 2021,
  journal = {Proc. Combust. Inst.},
  volume = {38},
  number = {2},
  pages = {2691--2699},
  issn = {15407489},
  urldate = {2023-03-15},
  langid = {english}
}

@article{domingo2008,
  title = {Large-Eddy Simulation of a Lifted Methane Jet Flame in a Vitiated Coflow},
  author = {Domingo, P. and Vervisch, L. and Veynante, D.},
  year = 2008,
  month = feb,
  journal = {Combust. Flame},
  volume = {152},
  number = {3},
  pages = {415--432},
  urldate = {2022-11-26},
  langid = {english}
}

@article{eberle2018,
  title = {Toward Finite-Rate Chemistry Large-Eddy Simulations of Sooting Swirl Flames},
  author = {Eberle, Christian and Gerlinger, Peter and Geigle, Klaus Peter and Aigner, Manfred},
  year = 2018,
  month = jul,
  journal = {Combust. Sci. Technol.},
  volume = {190},
  number = {7},
  pages = {1194--1217},
  publisher = {Taylor \& Francis},
  issn = {0010-2202}
}

@article{elhelou2021,
  title = {Experimental {{Investigation}} of {{Soot Production}} and {{Oxidation}} in a {{Lab-Scale Rich}}--{{Quench}}--{{Lean}} ({{RQL}}) {{Burner}}},
  author = {El Helou, Ingrid and Skiba, Aaron W. and Mastorakos, Epaminondas},
  year = 2021,
  month = apr,
  journal = {Flow Turbul. Combust.},
  volume = {106},
  number = {4},
  pages = {1019--1041},
  urldate = {2023-03-15},
  langid = {english}
}

@article{felden2018,
  title = {Impact of Direct Integration of {{Analytically Reduced Chemistry}} in {{LES}} of a Sooting Swirled Non-Premixed Combustor},
  author = {Felden, Anne and Riber, Eleonore and Cuenot, Benedicte},
  year = 2018,
  month = may,
  journal = {Combust. Flame},
  volume = {191},
  pages = {270--286},
  urldate = {2024-05-06},
  langid = {english}
}

@article{ferraro2022,
  title = {Soot Particle Size Distribution Reconstruction in a Turbulent Sooting Flame with the Split-Based Extended Quadrature Method of Moments},
  author = {Ferraro, Federica and Gierth, Sandro and Salenbauch, Steffen and Han, Wang and Hasse, Christian},
  year = 2022,
  month = jul,
  journal = {Phys. Fluids},
  volume = {34},
  number = {7},
  pages = {075121},
  urldate = {2023-03-16},
  langid = {english}
}

@article{franzelli2023,
  title = {Assessment of {{LES}} of Intermittent Soot Production in an Aero-Engine Model Combustor Using High-Speed Measurements},
  author = {Franzelli, B. and Tardelli, L. and St{\"o}hr, M. and Geigle, K.P. and Domingo, P.},
  year = 2023,
  journal = {Proc. Combust. Inst.},
  volume = {39},
  number = {4},
  pages = {4821--4829},
  issn = {15407489},
  urldate = {2024-04-25},
  langid = {english}
}

@article{Frenklach.1991,
  title = {Detailed {{Modeling}} of {{Soot Particle Nucleation}} and {{Growth}}},
  author = {Frenklach, Michael and Wang, Hai},
  year = 1991,
  month = jan,
  journal = {Proc. Combust. Inst.},
  volume = {23},
  number = {1},
  pages = {1559--1566}
}

@book{friendlander2000smoke,
  title = {Smoke, Dust and Haze: {{Fundamentals}} of Aerosol Dynamics},
  author = {Friendlander, {\relax SK}},
  year = 2000,
  publisher = {Oxford University Press}
}

@article{garcia-oliver2024,
  title = {{{LES}} of a Pressurized Sooting Aero-Engine Model Burner Using a Computationally Efficient Discrete Sectional Method Coupled to Tabulated Chemistry},
  author = {{Garc{\'i}a-Oliver}, J.M. and Pastor, J.M. and Olmeda, I. and Kalbhor, A. and Mira, D. and {van Oijen}, J.A.},
  year = 2024,
  month = feb,
  journal = {Combust. Flame},
  volume = {260},
  pages = {113198},
  urldate = {2023-12-11},
  langid = {english}
}

@article{Geigle.2015,
  title = {Investigation of Soot Formation in Pressurized Swirl Flames by Laser Measurements of Temperature, Flame Structures and Soot Concentrations},
  author = {Geigle, Klaus Peter and K{\"o}hler, Markus and O'Loughlin, William and Meier, Wolfgang},
  year = 2015,
  journal = {Proc. Combust. Inst.},
  volume = {35},
  number = {3},
  pages = {3373--3380},
  issn = {15407489},
  urldate = {2024-03-06},
  langid = {english}
}

@inproceedings{Giusti.2018,
  title = {Numerical {{Investigation}} of {{Flame Structure}} and {{Soot Formation}} in a {{Lab-Scale Rich-Quench-Lean Burner}}},
  booktitle = {{{GT2018-76705 V04BT04A032}}},
  author = {Giusti, Andrea and Gkantonas, Savvas and Foale, Jenna M. and Mastorakos, Epaminondas},
  year = 2018,
  month = jun,
  publisher = {ASME},
  address = {Oslo},
  urldate = {2023-03-15},
  langid = {english}
}

@article{gkantonas2020,
  title = {Comprehensive Soot Particle Size Distribution Modelling of a Model {{Rich-Quench-Lean}} Burner},
  author = {Gkantonas, Savvas and Sirignano, Mariano and Giusti, Andrea and D'Anna, Andrea and Mastorakos, Epaminondas},
  year = 2020,
  month = jun,
  journal = {Fuel},
  volume = {270},
  pages = {117483},
  urldate = {2023-03-15},
  langid = {english}
}

@article{govert2015,
  title = {Turbulent Combustion Modelling of a Confined Premixed Jet Flame Including Heat Loss Effects Using Tabulated Chemistry},
  author = {G{\"o}vert, S. and Mira, D. and Kok, J.B.W. and V{\'a}zquez, M. and Houzeaux, G.},
  year = 2015,
  month = oct,
  journal = {Appl. Energy},
  volume = {156},
  pages = {804--815},
  issn = {03062619},
  urldate = {2025-05-05},
  langid = {english}
}

@article{govert2017,
  title = {Heat Loss Prediction of a Confined Premixed Jet Flame Using a Conjugate Heat Transfer Approach},
  author = {G{\"o}vert, S. and Mira, D. and {Zavala-Ake}, M. and Kok, J.B.W. and V{\'a}zquez, M. and Houzeaux, G.},
  year = 2017,
  month = apr,
  journal = {Int. J. Heat Mass Transf.},
  volume = {107},
  pages = {882--894},
  issn = {00179310},
  urldate = {2025-05-05},
  langid = {english}
}

@article{govert2018,
  title = {The {{Effect}} of {{Partial Premixing}} and {{Heat Loss}} on the {{Reacting Flow Field Prediction}} of a {{Swirl Stabilized Gas Turbine Model Combustor}}},
  author = {G{\"o}vert, S. and Mira, D. and Kok, J. B. W. and V{\'a}zquez, M. and Houzeaux, G.},
  year = 2018,
  month = mar,
  journal = {Flow Turbul. Combust.},
  volume = {100},
  number = {2},
  pages = {503--534},
  issn = {1386-6184, 1573-1987},
  urldate = {2022-01-13},
  langid = {english}
}

@inproceedings{Grader.2018,
  title = {{{LES}} of a {{Pressurized}}, {{Sooting Aero-Engine Model Combustor}} at {{Different Equivalence Ratios With}} a {{Sectional Approach}} for {{PAHs}} and {{Soot}}},
  booktitle = {{{GT2018-75254 V04AT04A012}}},
  author = {Grader, Martin and Eberle, Christian and Gerlinger, Peter and Aigner, Manfred},
  year = 2018,
  month = jun,
  publisher = {ASME},
  address = {Oslo},
  urldate = {2024-06-05}
}

@phdthesis{helou2021,
  title = {Experimental {{Investigation}} of {{Large Scale Aerodynamics}} on {{Soot Emissions}} in {{Swirling Flows}}},
  author = {Helou, Ingrid El},
  year = 2021,
  address = {Cambridge, United Kingdom},
  langid = {english},
  school = {University of Cambridge}
}

@article{Hoerlle.2019,
  title = {Effects of {{CO2}} Addition on Soot Formation of Ethylene Non-Premixed Flames under Oxygen Enriched Atmospheres},
  author = {Hoerlle, Cristian A. and Pereira, Fernando M.},
  year = 2019,
  journal = {Combust. Flame},
  volume = {203},
  pages = {407--423}
}

@article{ihme2008,
  title = {Prediction of Extinction and Reignition in Nonpremixed Turbulent Flames Using a Flamelet/Progress Variable Model: 2. {{Application}} in {{LES}} of {{Sandia}} Flames {{D}} and {{E}}},
  author = {Ihme, M. and Pitsch, H.},
  year = 2008,
  month = oct,
  journal = {Combust. Flame},
  volume = {155},
  number = {1},
  pages = {90--107}
}

@article{kalbhor2020,
  title = {Effects of Hydrogen Enrichment and Water Vapour Dilution on Soot Formation in Laminar Ethylene Counterflow Flames},
  author = {Kalbhor, Abhijit and {van Oijen}, Jeroen},
  year = 2020,
  month = sep,
  journal = {Int. J. Hydrogen Energy},
  volume = {45},
  number = {43},
  pages = {23653--23673},
  urldate = {2023-11-21},
  langid = {english}
}

@article{kalbhor2021,
  title = {An Assessment of the Sectional Soot Model and {{FGM}} Tabulated Chemistry Coupling in Laminar Flame Simulations},
  author = {Kalbhor, Abhijit and van Oijen, Jeroen},
  year = 2021,
  month = jul,
  journal = {Combust. Flame},
  volume = {229},
  pages = {111381},
  urldate = {2023-03-31},
  langid = {english}
}

@article{kalbhor2023,
  title = {A Computationally Efficient Approach for Soot Modeling with Discrete Sectional Method and {{FGM}} Chemistry},
  author = {Kalbhor, Abhijit and Mira, Daniel and {van Oijen}, Jeroen},
  year = 2023,
  month = sep,
  journal = {Combust. Flame},
  volume = {255},
  pages = {112868},
  urldate = {2023-06-09},
  langid = {english}
}

@article{kalbhor2024,
  title = {{{LES}} Investigation of Soot Formation in a Turbulent Non-Premixed Jet Flame with Sectional Method and {{FGM}} Chemistry},
  author = {Kalbhor, Abhijit and Mira, Daniel and Both, Ambrus and {van Oijen}, Jeroen},
  year = 2024,
  month = jan,
  journal = {Combust. Flame},
  volume = {259},
  pages = {113128},
  urldate = {2023-11-16},
  langid = {english}
}

@article{kempf2005,
  title = {Efficient {{Generation}} of {{Initial-}} and {{Inflow-Conditions}} for {{Transient Turbulent Flows}} in {{Arbitrary Geometries}}},
  author = {Kempf, A. and Klein, M. and Janicka, J.},
  year = 2005,
  month = jan,
  journal = {Flow Turbul. Combust.},
  volume = {74},
  number = {1},
  pages = {67--84},
  urldate = {2023-02-23},
  langid = {english}
}

@article{kumar1996,
  title = {On the Solution of Population Balance Equations by Discretization---{{I}}. {{A}} Fixed Pivot Technique},
  author = {Kumar, Sanjeev and Ramkrishna, D.},
  year = 1996,
  month = apr,
  journal = {Chem. Eng. Sci.},
  volume = {51},
  number = {8},
  pages = {1311--1332},
  urldate = {2023-11-21},
  langid = {english}
}

@article{li2024,
  title = {Microphysical Properties of Atmospheric Soot and Organic Particles: Measurements, Modeling, and Impacts},
  shorttitle = {Microphysical Properties of Atmospheric Soot and Organic Particles},
  author = {Li, Weijun and Riemer, Nicole and Xu, Liang and Wang, Yuanyuan and Adachi, Kouji and Shi, Zongbo and Zhang, Daizhou and Zheng, Zhonghua and Laskin, Alexander},
  year = 2024,
  month = mar,
  journal = {npj Clim. Atmos. Sci.},
  volume = {7},
  number = {1},
  pages = {65},
  issn = {2397-3722},
  urldate = {2025-04-30},
  langid = {english}
}

@article{massey2021,
  title = {Modelling {{Heat Loss Effects}} in the {{Large Eddy Simulation}} of a {{Lean Swirl-Stabilised Flame}}},
  author = {Massey, James C. and Chen, Zhi X. and Swaminathan, Nedunchezhian},
  year = 2021,
  month = apr,
  journal = {Flow Turbul. Combust.},
  volume = {106},
  number = {4},
  pages = {1355--1378},
  urldate = {2023-11-20},
  langid = {english}
}

@article{mira2020,
  title = {Numerical {{Characterization}} of a {{Premixed Hydrogen Flame Under Conditions Close}} to {{Flashback}}},
  author = {Mira, D. and Lehmkuhl, O. and Both, A. and Stathopoulos, P. and Tanneberger, T. and Reichel, T. G. and Paschereit, C. O. and V{\'a}zquez, M. and Houzeaux, G.},
  year = 2020,
  month = mar,
  journal = {Flow Turbul. Combust.},
  volume = {104},
  number = {2-3},
  pages = {479--507},
  issn = {1386-6184, 1573-1987},
  urldate = {2024-01-04},
  langid = {english}
}

@article{miramartinez2014,
  title = {Numerical Assessment of Subgrid Scale Models for Scalar Transport in Large-Eddy Simulations of Hydrogen-Enriched Fuels},
  author = {Mira Martinez, D. and Jiang, X. and Moulinec, C. and Emerson, D.R.},
  year = 2014,
  month = may,
  journal = {Int. J. Hydrogen Energy},
  volume = {39},
  number = {14},
  pages = {7173--7189},
  urldate = {2022-11-23},
  langid = {english}
}

@article{Neoh.1985,
  title = {Effect of Oxidation on the Physical Structure of Soot},
  author = {Neoh, K.G. and Howard, J.B. and Sarofim, A.F.},
  year = 1985,
  month = jan,
  journal = {Proc. Combust. Inst.},
  volume = {20},
  number = {1},
  pages = {951--957},
  issn = {0082-0784},
  doi = {10.1016/S0082-0784(85)80584-1}
}

@article{pachano2024,
  title = {Analysis of Soot Formation in a Lab-Scale {{Rich-Quench-Lean}} Combustor Using {{LES}} with Tabulated Chemistry},
  author = {Pachano, Leonardo and Kalbhor, Abhijit and Mira, Daniel and Van Oijen, Jeroen},
  year = 2024,
  journal = {Proc. Combust. Inst.},
  volume = {40},
  number = {1-4},
  pages = {105451},
  issn = {15407489},
  urldate = {2025-02-05},
  langid = {english}
}

@book{poinsot2005,
  title = {Theoretical and Numerical Combustion},
  author = {Poinsot, T. and Veynante, D.},
  year = 2005,
  publisher = {RT Edwards, Inc.}
}

@phdthesis{somers1994,
  title = {The Simulation of Flat Flames with Detailed and Reduced Chemical Models},
  author = {Somers, L},
  year = 1994,
  address = {Eindhoven, the Netherlands},
  school = {Eindhoven University of Technology}
}

@article{thomson2023,
  title = {Modeling Soot Formation in Flames and Reactors: {{Recent}} Progress and Current Challenges},
  author = {Thomson, Murray J.},
  year = 2023,
  journal = {Proc. Combust. Inst.},
  volume = {39},
  number = {1},
  pages = {805--823},
  urldate = {2023-11-23},
  langid = {english}
}

@article{vanoijen2016,
  title = {State-of-the-Art in Premixed Combustion Modeling Using Flamelet Generated Manifolds},
  author = {{van Oijen}, J.A. and Donini, A. and Bastiaans, R.J.M. and {ten Thije Boonkkamp}, J.H.M. and {de Goey}, L.P.H.},
  year = 2016,
  month = nov,
  journal = {Prog. Energy Combust. Sci.},
  volume = {57},
  pages = {30--74},
  urldate = {2023-01-04},
  langid = {english}
}

@article{vazquez2016,
  title = {Alya: {{Multiphysics}} Engineering Simulation toward Exascale},
  author = {V{\'a}zquez, M. and Houzeaux, G. and Koric, S. and Artigues, A. and {Aguado-Sierra}, Jazmin and Ar{\'i}s, R. and Mira, D. and Calmet, H. and Cucchietti, F. and Owen, H. and Taha, A. and Burness, E. and Cela, J. M. and Valero, M.},
  year = 2016,
  month = may,
  journal = {J. Comput. Sci.},
  volume = {14},
  pages = {15--27},
  urldate = {2022-02-25},
  langid = {english}
}

@article{vreman2004,
  title = {An Eddy-Viscosity Subgrid-Scale Model for Turbulent Shear Flow: {{Algebraic}} Theory and Applications},
  author = {Vreman, A. W.},
  year = 2004,
  month = oct,
  journal = {Phys. Fluids},
  volume = {16},
  number = {10},
  pages = {3670--3681},
  urldate = {2022-11-23},
  langid = {english}
}

@article{Wang.2013,
  title = {A {{PAH}} Growth Mechanism and Synergistic Effect on {{PAH}} Formation in Counterflow Diffusion Flames},
  author = {Wang, Yu and Raj, Abhijeet and Chung, Suk Ho},
  year = 2013,
  journal = {Combust. Flame},
  volume = {160},
  number = {9},
  pages = {1667--1676}
}

\end{document}